\documentclass[aip,jcp,amsmath,amssymb,floatfix,citeautoscript,reprint]{revtex4-2}

\usepackage{graphicx}
\usepackage{dcolumn}
\usepackage{bm}

\usepackage{braket}
\usepackage[usenames,dvipsnames]{color}
\usepackage[version=3]{mhchem}
\usepackage{ragged2e}
\usepackage[labelsep=period, format=plain, justification=justified,font=footnotesize]{caption}
\usepackage{times}
\usepackage{txfonts}
\usepackage{relsize}

\usepackage{xr}
\makeatletter
\newcommand*{\addFileDependency}[1]{
  \typeout{(#1)}
  \@addtofilelist{#1}
  \IfFileExists{#1}{}{\typeout{No file #1.}}
}
\makeatother
\newcommand*{\myexternaldocument}[1]{
    \externaldocument{#1}
    \addFileDependency{#1.tex}
    \addFileDependency{#1.aux}
}
\myexternaldocument{build/si}

\DeclareCaptionJustification{myjust}{\justifying}
\newcommand{\RNum}[1]{\uppercase\expandafter{\romannumeral #1\relax}}

\begin{document}

\title{Elucidating the role of hydrogen bonding in the optical spectroscopy of the solvated green fluorescent protein chromophore: using machine learning to establish the importance of high-level electronic structure}

\author{Michael S. Chen}
\altaffiliation{These authors contributed equally to this work.}
\affiliation{Department of Chemistry, Stanford University, Stanford, California 94305, USA}

\author{Yuezhi Mao}
\altaffiliation{These authors contributed equally to this work.}
\affiliation{Department of Chemistry, Stanford University, Stanford, California 94305, USA}

\author{Andrew Snider}
\affiliation{Chemistry and  Biochemistry, University of California Merced, Merced, California 95343, USA}

\author{Prachi Gupta}
\affiliation{Chemistry and Biochemistry, University of California Merced, Merced, California 95343, USA}

\author{Andr\'es Montoya-Castillo}
\affiliation{Department of Chemistry, University of Colorado, Boulder, Boulder, Colorado 80309, USA}

\author{Tim J. Zuehlsdorff}
 \affiliation{Department of Chemistry, Oregon State University, Corvallis, Oregon 97331, USA}

\author{Christine M. Isborn}
\email{cisborn@ucmerced.edu}
\affiliation{Chemistry and Biochemistry, University of California Merced, Merced, California 95343, USA}

\author{Thomas E. Markland}
\email{tmarkland@stanford.edu}
\affiliation{Department of Chemistry, Stanford University, Stanford, California 94305, USA}

\date{\today}

\begin{abstract}
Hydrogen bonding interactions with chromophores in chemical and biological environments play a key role in determining their electronic absorption and relaxation processes, which are manifested in their linear and multidimensional optical spectra. For chromophores in the condensed phase, the large number of atoms needed to simulate the environment has traditionally prohibited the use of high-level excited-state electronic structure methods. By leveraging transfer learning, we show how to construct machine-learned models to accurately predict high-level excitation energies of a chromophore in solution from only 400 high-level calculations. We show that when the electronic excitations of the green fluorescent protein chromophore in water are treated using EOM-CCSD embedded in a DFT description of the solvent, the optical spectrum is correctly captured and that this improvement arises from correctly treating the coupling of the electronic transition to electric fields, which leads to a larger response upon hydrogen bonding between the chromophore and water. 
\end{abstract}

\maketitle
\normalsize

Chromophores' functions in biological and technological applications, ranging from photosynthesis to fluorescence imaging\cite{Scholes2011lessons,SchlauCohen2015principles,Wang2008fluorescence}, benefit immensely from the tunable nature of their photophysical properties. Owing to these important uses, a wealth of previous studies have focused on characterizing how different chromophores' excitations are tuned by the interactions with their chemical environment\cite{Su1991time,DiBella1994environmental,Bell2000probing,Schellenberg2001resonance,Dong2006solvatochromism,Naumov2010topochemistry,Day2009fluorescent,Oltrogge2015short,Lin2019unified,Lin2020unusual,Lin2022energetic} in order to uncover the physical basis underlying these phenomena and guide the design of chromophores with particular target properties. Simulations of chromophores in protein scaffolds or in solution have also provided extensive insights into the coupling between the chemical environment and how the interplay of the electronic and nuclear dynamics give rise to their linear and multidimensional optical spectra\cite{Filippi_2009,Rivalta2014ab,Davari_2014,Huang2021ab,Zuehlsdorff2020influence,Lu2021influence,Zuehlsdorff2021vibronic,Abou2021modeling}. However, for chromophores in solvated condensed phase environments, performing the required excited state electronic structure calculations with a quantum treatment of the environment is extremely computationally costly owing to the large system sizes (typically 300-500 atoms) needed to treat the chromophore and its environment. For condensed-phase systems, this electronic structure computational cost is combined with the need to perform many thousands of electronic structure calculations to account for the many thermally relevant configurations of the chromophore and solvent. Due to these challenges, the vast majority of previous studies that sample chromophore-environment excitation energies have used time-dependent density functional theory (TDDFT) \cite{Casida1995,Dreuw2005single} to treat the electronic excitations since it serves as a practical compromise between computational efficiency and accuracy. However, although for a range of chromophores in solution TDDFT has been shown to produce accurate linear spectra\cite{Ge2015accurate, Hiyama2017effect,GarciaIriepa2018simulation,Srsen2020, Manian20222, Gomez2023}, in particular when combined with approaches that account for vibronic effects that are not present in ensemble methods\cite{Zuehlsdorff2019optical,Zuehlsdorff2021vibronic,Cerezo2020,Cerezo2023, Petrusevich2023}, challenging cases remain where a higher level treatment of the electronic structure is essential. 

\begin{figure*}[]
    \begin{center}  \includegraphics[width=0.96\textwidth]{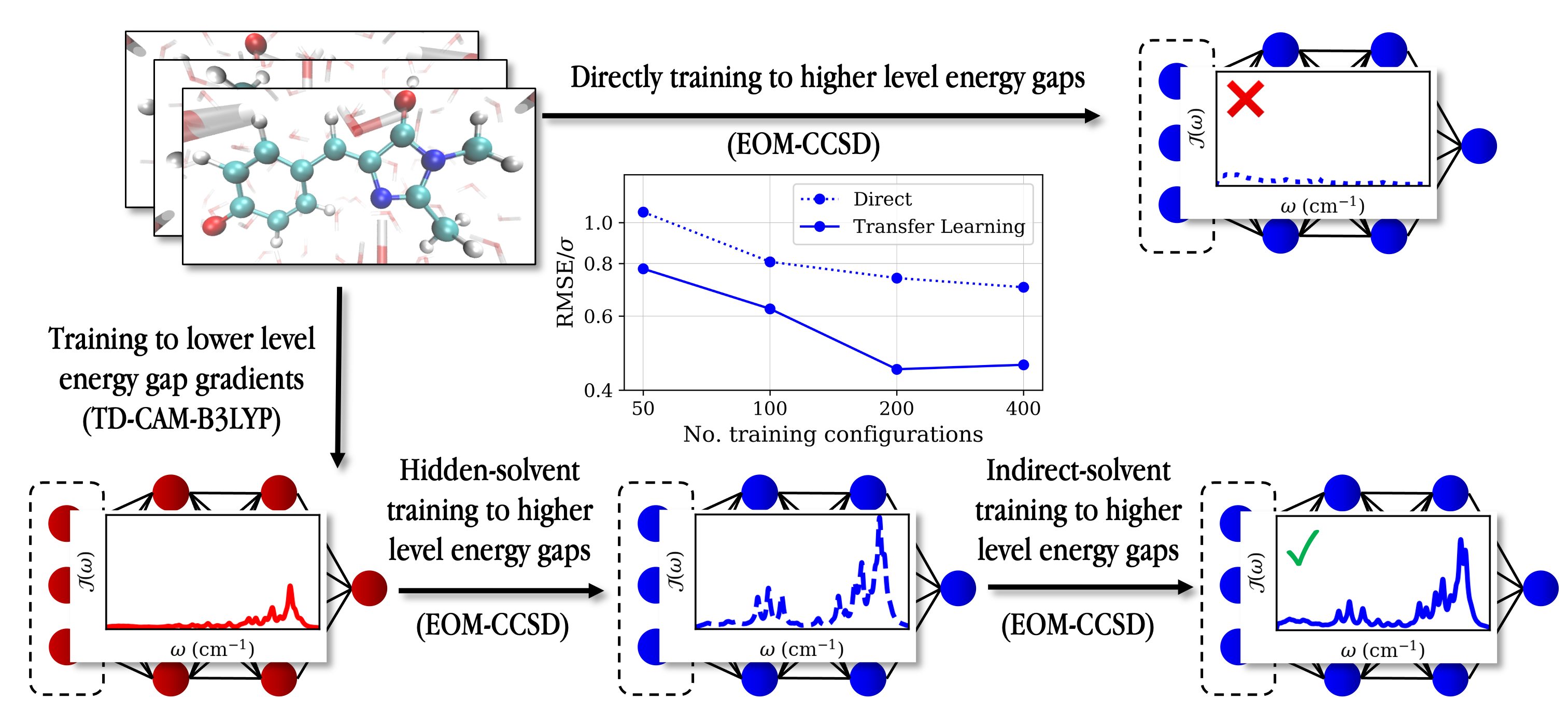}
    \end{center}
    \caption{An overview of our transfer learning procedure and how it performs compared to a direct fit of the EOM-CCSD energy gaps for the GFP chromophore in water. First we construct an ML model to predict TD-CAM-B3LYP energy gaps by training on the gradients of the electronic energy gap, i.e., the excited state gradient minus that in the ground state, from 300 configurations. We then use that fit to initialize a transfer learning fit to the EOM-CCSD energy gaps for 400 configurations, first employing a hidden-solvent ML approach that only accounts for the chromophore atom positions before incorporating solvent atom positions with a subsequent indirect-solvent ML model using the same 400 EOM-CCSD energy gaps. This transfer learning model is considerably more accurate (lower RMSE/$\sigma$) than a direct model fit to the same number of EOM-CCSD energy gaps and accurately captures the intensity of the peaks in the spectral density $\mathcal{J}(\omega)$.}
	\label{fig:summary_schematic}
\end{figure*}

One system that poses a significant challenge to simulation is the anionic GFP chromophore (\textit{p}-hydroxybenzylidene-2,3-dimethylimidazolinone, HBDI$^-$) in water. The optical absorption spectrum of the anionic GFP chromophore in aqueous solution is much broader than observed in the protein or vacuum, and the full extent of this experimentally observed broadening has not been captured in the many previous simulation studies of the GFP chromophore anion in water~\cite{Avila_Ferrer_2014,Zuehlsdorff2018combining, Zuehlsdorff2018unraveling, Raucci_2020}. The origins of the overly narrow optical absorption peak have been previously suggested to arise from not accounting for the thermal distribution of non-planar chromophore geometries~\cite{Avila_Ferrer_2014} and neglecting vibronic~\cite{Raucci_2020} effects. However, in our recent study~\cite{Zuehlsdorff2018unraveling} we included both of these effects, along with the nuclear quantum effects (NQEs) that are known to provide additional spectral broadening and red shift optical absorption spectra,\cite{Svoboda2011simulations,Law2015role,Sappati2016nuclear,Grisanti2017computational,Law2018importance,Feher2021multiscale} by combining \textit{ab initio} path integral simulations, which treat the quantum mechanical nature of the nuclei,\cite{Markland2018nuclear} with a combined ensemble-Franck-Condon method\cite{Zuehlsdorff2018combining}, which accounts for vibronic effects, explicit solvent effects, and non-planar chromophore geometries. Even with all of these effects included our spectral simulation, the breadth of the absorption spectrum of the GFP chromophore in water was still underestimated when TDDFT was used to describe the electronic excitations. These studies suggest that to correctly describe the GFP chromophore in water one may need to go beyond a TDDFT description of its electronic structure.

Equation-of-motion coupled-cluster with singles and doubles (EOM-CCSD)~\cite{Stanton1993equation, Comeau1993equation, Krylov2008equation} has been established as an accurate high-level method for calculating electronic excited states,\cite{Loos2018mountaineering,Loos2020mountaineering} and has been employed for calculating the excited states of the GFP chromophore.\cite{Filippi_2009, Polyakov2010} However, because of its $\mathcal{O}(N^6)$ system-size scaling it is not currently practical to compute the electronic excitations of the GFP chromophore with full explicit solvation, a system comprising at least $\sim$350 atoms, all at the EOM-CCSD level. To avoid the cost of a full EOM-CCSD calculation on the entire system, one can leverage a recently introduced projection-based embedding theory~\cite{Manby2012simple, Lee2019projection, Bennie2017pushing, Claudino2019automatic, Claudino2019simple, Parravicini2021embedded} to embed an EOM-CCSD description of the chromophore within an explicit solvation environment described by ground-state DFT. Using our implementation of embedded EOM-CCSD in Q-Chem \cite{Epifanovsky2021} makes it practical to achieve 100's of excited state energies for chromophores, such as the anionic GFP chromophore, in solution. 

However, when dynamical approaches\cite{Zuehlsdorff2019optical}, which allow for the accurate accounting of anharmonic and vibronic effects within an explicit solvent environment, are used to simulate linear and multidimensional optical spectra, tens of thousands of excited-state energy calculations are required to compute the necessary time correlation functions, thus presenting a massive computational challenge. To address this challenge, here we introduce a data-efficient transfer learning protocol to produce accurate machine-learning (ML) models of the high-level excited-state electronic structure for chromophores in the condensed phase by training on only hundreds of high-level excited-state energy calculations. We utilize a hierarchical approach where initially the model is imbued with the fundamental physics of a low-level electronic structure approach for which generating many 1000's of excited state calculations is straightforward. We then show how this initial low-level ML model that qualitatively captures the character of the electronic excitations, such as which nuclear motions couple most strongly to the electronic excitation, can be enhanced to achieve quantitative accuracy by transfer learning using only 100's of high-level energies. The data efficiency of this approach allows us to train ML models to high-level electronic structure with the solvation environment explicitly included, expanding on other recent studies which have trained ML models to high-level electronic structure calculations in the gas phase\cite{Chen2018deep,Westermayr2020deep} or with implicit solvation\cite{Chen2022UV, Chen2023} or to lower-level electronic structure methods such as TDDFT. \cite{Hase2016machine,Liu2017direct,Pronobis2018capturing,Ye2019neural,Chen2020exploiting,Xue2020machine,Farahvash2020machine,Bononi2020bathochromic,Lu2020deep,Hullar2020photodecay,Hullar2022atmospheric,Petrusevich2023,Cignoni2023machine,Dral2021molecular,Westermayr2021} In particular, here we demonstrate that one can develop an accurate ML model of electronic excitations at the level of EOM-CCSD embedded in DFT by starting from either a TDDFT or configuration interaction singles (CIS) \cite{Bene1971self,Foresman1992toward} low-level treatment of the excited-state electronic structure using only 400 embedded EOM-CCSD energy calculations.

Using our ML model trained to high-level embedded EOM-CCSD calculations in conjunction with a dynamical approach to computing the linear and multidimensional absorption spectra, which leverages the second order cumulant expansion of the energy gap operator\cite{Mukamel1995,Zuehlsdorff2019optical}, we show that one can obtain quantitative agreement with the shape of the linear optical absorption spectrum of the anionic GFP chromophore in water and that by also including NQEs one can also accurately capture the position of the peak. We show that the ability of embedded EOM-CCSD to capture the shape of the spectrum more accurately than TDDFT arises from a different physical picture of how hydrogen-bonding and, more generally, how solvent electric fields couple to the electronic excitations of the anionic GFP chromophore in water.

Figure~\ref{fig:summary_schematic} provides an overview of our strategy to develop a data-efficient ML model to accurately predict energy gaps at the level of embedded EOM-CCSD for the GFP chromophore in water. This approach harnesses the ability to efficiently generate TDDFT (or CIS) electronic excited-state energies and gradients for chromophores in solution using GPU accelerated electronic structure calculations~\cite{Isborn2011excited,Seritan_TeraChem_2021}, which allow 1,000's-10,000's of calculations to be performed. In contrast, embedded EOM-CCSD calculations come at a substantially higher computational cost and do not yet have gradients implemented, thereby limiting one to only 100's of excited-state energies without their corresponding gradients. Given this small amount of high-level EOM-CCSD data, the top part of Fig.~\ref{fig:summary_schematic} shows the futility of directly training to 400 excitation energies. The resulting ML model of the high-level data performs poorly on the validation set with a root-mean-square error (RMSE) of 0.21~eV, which is close to the standard deviation of the energy gaps that the model was fit to ($\sigma=0.30$~eV). Figure~\ref{fig:summary_schematic} shows the spectral density, $\mathcal{J}(\omega)$, which encodes the coupling of the vibrational modes of the system to the electronic excitations, obtained from the ML model that was directly fit to the high-level electronic energy gaps. Other than a small amount of intensity in the low-frequency region, the resulting ML-predicted spectral density fails to capture the vibronic coupling between higher-frequency vibrations and the electronic excitation. The lower part of Fig.~\ref{fig:summary_schematic} shows our transfer learning protocol for training a data-efficient model for excited state electronic energy gaps. This procedure involves three stages:
\begin{enumerate}
    \item We first train an ML model on a low-level excited-state electronic structure approach where it is computationally feasible to generate a large training set to provide a converged model at that level of theory.
    \item This ML model is then used to initialize the training of a ``hidden-solvent" model\cite{Chen2020exploiting} to the small dataset of higher-level calculations. In the hidden-solvent model, the ML model is trained to the solvated energy gap while ignoring solvent atom positions giving an effective mean-field model of the excitations.
    \item The hidden-solvent ML model is then used to initialize the training of an ``indirect-solvent"\cite{Chen2020exploiting} ML model, where solvent atom positions are now incorporated to the descriptors,  using the small dataset of higher-level calculations.
\end{enumerate}
This hierarchical protocol exploits the readily available low-level data and combines it with judicious use of the high-level data. In particular, our protocol exploits the fact that, although the low-level data may not give a quantitative description of exactly how nuclear motions of the chromophore and solvent couple to the electronic transition, it does encode qualitative information about which atoms couple to the transition. This can be seen by comparing the bottom left and bottom right plots of Fig.~\ref{fig:summary_schematic} where the low-level (TD-CAM-B3LYP) ML model obtains a spectral density containing peaks in many of the same positions as high-level (embedded EOM-CCSD) but with different intensities. The second stage of the transfer learning procedure concentrates the small amount of high-level data on tuning the parameters of the low-level model with a chromophore-centric focus since the electronic excitation is primarily localized on the chromophore. The third stage provides a final tuning to the ML model by incorporating how the solvent atoms' positions modulate the electronic excitation beyond the mean-field description engendered in the hidden solvent model.

In practice, we performed our transfer learning procedure using two different low-level electronic structure methods, TDDFT (with the CAM-B3LYP \cite{Yanai2004new} functional) and CIS, and in both cases targeted embedded EOM-CCSD as the high-level method. The low-level ML models were trained to the gradients of the electronic energy gap obtained from 300 solvated chromophore configurations (see SI Sec.~\ref{si:sec:grad-training}). We note that despite this seemingly small number of configurations the ability to train on the gradients of the electronic energy gap provides extensive data since there are 3$N$ (where $N$ is the number of atoms in the simulation, which is 528 in our case) gradients per configuration. The high-level hidden-solvent and indirect-solvent ML models were trained on 400 embedded EOM-CCSD electronic energy gaps.

\begin{figure}[]
    \begin{center}
        \includegraphics[width=0.48\textwidth]{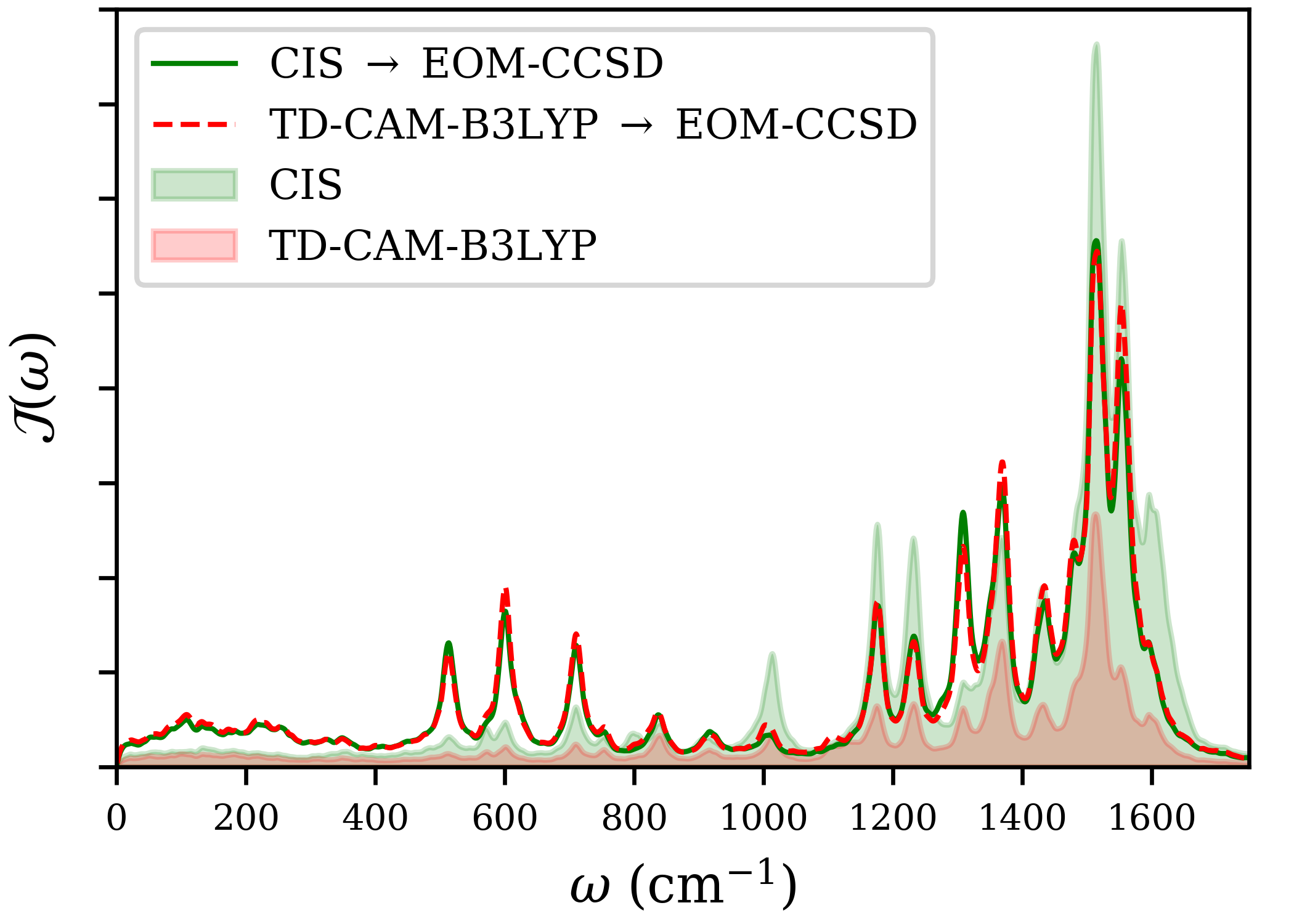}
    \end{center}
    \caption{Demonstration of the consistent results obtained from our transfer learning protocol when starting from two different low-level electronic structure methods. Shown are the spectral densities obtained from our embedded EOM-CCSD transfer-learned ML models starting from two different low-level methods: CIS (green) and TD-CAM-B3LYP (red). The two embedded EOM-CCSD models give consistent spectral densities (red and green lines) despite the two initial low-level methods giving distinctly different spectral densities (red and green shaded regions).}
	\label{fig:tf_hysteresis}
\end{figure}

By employing our transfer learning protocol, the final indirect-solvent ML model trained to 400 embedded EOM-CCSD energy gaps is considerably more accurate than the directly trained model, with a validation set RMSE of 0.128~eV vs. 0.195~eV. The middle figure in Fig.~\ref{fig:summary_schematic} shows that we obtain a consistent accuracy improvement for all the training set sizes tested. The improved accuracy of the transfer learned ML model is also reflected in the richer array of features present in the corresponding spectral density (Fig.~\ref{fig:summary_schematic}, bottom right) that were completely missing from the spectral density obtained from the directly trained ML model (Fig.~\ref{fig:summary_schematic}, top right). Although both of these observations regarding the accuracy of the high-level ML model obtained from transfer learning are promising, since the model was initialized from TDDFT there is a distinct risk of it inheriting the defects in this electronic structure method rather than being a faithful representation of the EOM-CCSD result. To assess the role of hysteresis arising from the initial low-level model, we also performed our transfer learning procedure starting from CIS as the low-level method. As shown in Figure~\ref{fig:tf_hysteresis}, the spectral density obtained from CIS is distinctly different from that obtained from TDDFT, with CIS showing a significant increase in intensity in the high-frequency region that leads to strong vibronic features in the linear absorption spectrum. However, when the transfer learning procedure is applied the spectral densities obtained for EOM-CCSD starting from either CIS or TDDFT are in excellent agreement. This reaffirms that, even when using only 400 EOM-CCSD energies, our procedure can provide a model that avoids hysteresis effects from the low-level method and is thus capturing the electronic structure of the high-level method.

\begin{figure}[]
    \begin{center}
        \includegraphics[width=0.48\textwidth]{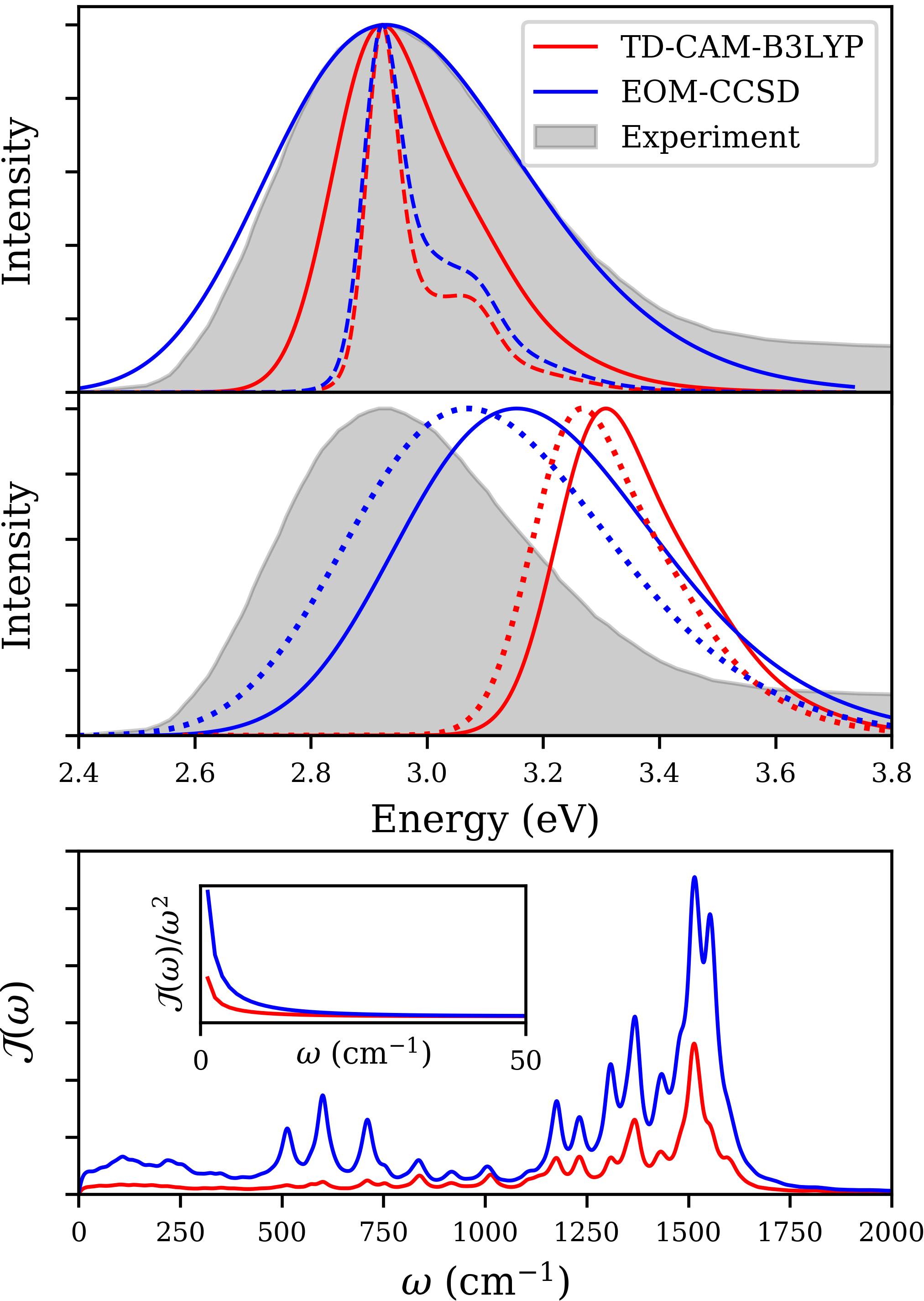}
    \end{center}
    \caption{Comparisons of the linear spectra (top, middle) and the spectral densities, $\mathcal{J}(\omega)$, (bottom) obtained from ML models of the electronic energy gaps to TD-CAM-B3LYP (red) or EOM-CCSD (blue). Top Panel: Linear absorption spectra for the GFP chromophore in aqueous solution with all spectra aligned with the experimental maximum (2.94 eV) to allow for comparison of the spectral shapes. The gas-phase linear spectra predicted by our gas-phase ML models are shown as dashed lines. Our EOM-CCSD model gives a linear electronic absorption spectrum for the GFP chromophore in water that much more accurately captures the experimental spectrum than TDDFT (TD-CAM-B3LYP). Middle Panel: Unshifted linear absorption spectra for TD-CAM-B3LYP (maximum at 3.30~eV) and EOM-CCSD (3.16~eV) for the GFP chromophore in water, with the corresponding spectra calculated using TRPMD depicted as dotted lines (maxima at 3.27~eV and 3.07~eV, respectively). Bottom Panel: The spectral density, $\mathcal{J}(\omega) $, as a function of frequency, $\omega$, obtained from the EOM-CCSD and TD-CAM-B3LYP models of the energy gaps. The inset shows $\mathcal{J}(\omega)/\omega^{2}$ at low frequencies since this is how the spectral density is weighted in the calculation of optical spectrum.}
	\label{fig:specdens_linspec}
\end{figure}

Upon validating our embedded EOM-CCSD ML model of the electronic excitations of the anionic GFP chromophore in water we can now assess the effect of using this higher-level electronic structure approach in capturing the optical absorption spectra. The top panel of Fig.~\ref{fig:specdens_linspec} compares the linear electronic absorption spectra of the GFP chromophore in the gas phase and in aqueous solution obtained from our TDDFT (TD-CAM-B3LYP) and EOM-CCSD ML models using a second order cumulant expansion of the electronic gap\cite{Zuehlsdorff2019optical} to obtain the absorption spectra from a ground state \textit{ab initio} molecular dynamics trajectory (see SI Secs.~\ref{si:sec:cumulant-approach} and \ref{si:sec:md-details})\cite{Zuehlsdorff2018unraveling}. In the top panel of Fig.~\ref{fig:specdens_linspec} all the spectra have been shifted to have their absorption maxima at the same excitation energy (matching that of the experimentally observed value of GFP in solution) to allow for comparison of the breadth and shape of their absorption peaks. The gas-phase absorption spectra (dashed lines) obtained from TD-CAM-B3LYP and EOM-CCSD have a very similar shape with a narrow main peak accompanied by a vibronic shoulder arising from the interaction of the electronic excitation with the intramolecular \ce{C=C} vibrations at $\sim$1500 cm$^{-1}$, with EOM-CCSD showing more intensity in the vibronic shoulder due to the larger intensity of these peaks in the spectral density. However, although the TD-CAM-B3LYP spectrum only slightly broadens (1.35 times the standard deviation) in aqueous solution, the EOM-CCSD spectrum more than doubles in width (2.14 times), in excellent agreement with the experimentally observed spectrum. Turning to the unshifted spectra, EOM-CCSD also improves the peak positions in both the gas (SI~Fig.~\ref{si:fig:linspec-unshifted}) and aqueous phases (middle panel of Fig.~\ref{fig:specdens_linspec}) with the peak maxima being 2.77~eV and 3.16~eV in the two respective phases compared to 2.56~eV and 2.94~eV\cite{Nielsen2001absorption} for experiment and 3.24~eV and 3.30~eV for TD-CAM-B3LYP. We note that the EOM-CCSD spectrum does not capture the high-energy tail of the experimental spectrum, which is likely attributable to contributions from higher-lying electronic excited states \cite{Zuehlsdorff2018unraveling} that we do not account for in this work where we focus solely on the S$_0 \rightarrow$ S$_1$ excitation of the GFP chromophore. These results highlight the failure of TD-CAM-B3LYP in capturing the width of the experimental linear electronic absorption spectrum for the GFP chromophore in water and the ability of higher-level methods like EOM-CCSD to better capture the experimental spectrum.

The origins of these differences can be explained by considering the spectral densities shown in the bottom panel of Fig.~\ref{fig:specdens_linspec} that give rise to the corresponding linear optical spectra. In the gas phase, EOM-CCSD and TD-CAM-B3LYP give almost identical spectra densities (see SI Fig.~\ref{si:fig:linspec-unshifted}), consistent with their very similar optical absorption spectra. However, in aqueous solution the EOM-CCSD spectral density exhibits higher intensities at all frequencies than obtained from TD-CAM-B3LYP, showing stronger coupling between nuclear motions and the electronic excitation when the higher-level method is used. To assess which of the regions of increased spectral density lead to the significant broadening of the spectrum of GFP in solution, in SI Fig.~\ref{si:fig:specdens-stitch} we show the effect on the predicted linear absorption spectrum of combining the high-frequency ($>$ 350~cm$^{-1}$) spectral density obtained with TDDFT with the low-frequency ($<$ 350~cm$^{-1}$) spectral density obtained from EOM-CCSD. This decomposition demonstrates that the origin of a substantial portion (40\%) of the peak broadening for GFP in aqueous solution upon going from TDDFT to EOM-CCSD arises from the increase in the intensity of the spectral density in the low-frequency ($<$ 350~cm$^{-1}$) region, with the remainder arising from the higher frequency regions (See SI Fig.~\ref{si:fig:specdens-stitch}). This result can also be rationalized by considering the connection between the spectral density and the optical absorption spectra; the linear spectrum is calculated by weighting the spectral density with a $1/\omega^2$ factor, and the inset in Fig.~\ref{fig:specdens_linspec} shows that differences in $\mathcal{J}(\omega)/\omega^2$ are more pronounced at lower frequencies. These lower frequency contributions are attributable to slower solvent-coupled modes\cite{Zuehlsdorff2020influence}, which are not present in the spectral density arising from either electronic structure method in the gas phase (see SI Sec.~\ref{si:fig:linspec-unshifted}), suggesting that the differences in the EOM-CCSD and TD-CAM-B3LYP spectra arise from differing treatments of the chromophore-solvent interactions for this system.

\begin{figure}[]
    \begin{center}
        \includegraphics[width=0.48\textwidth]{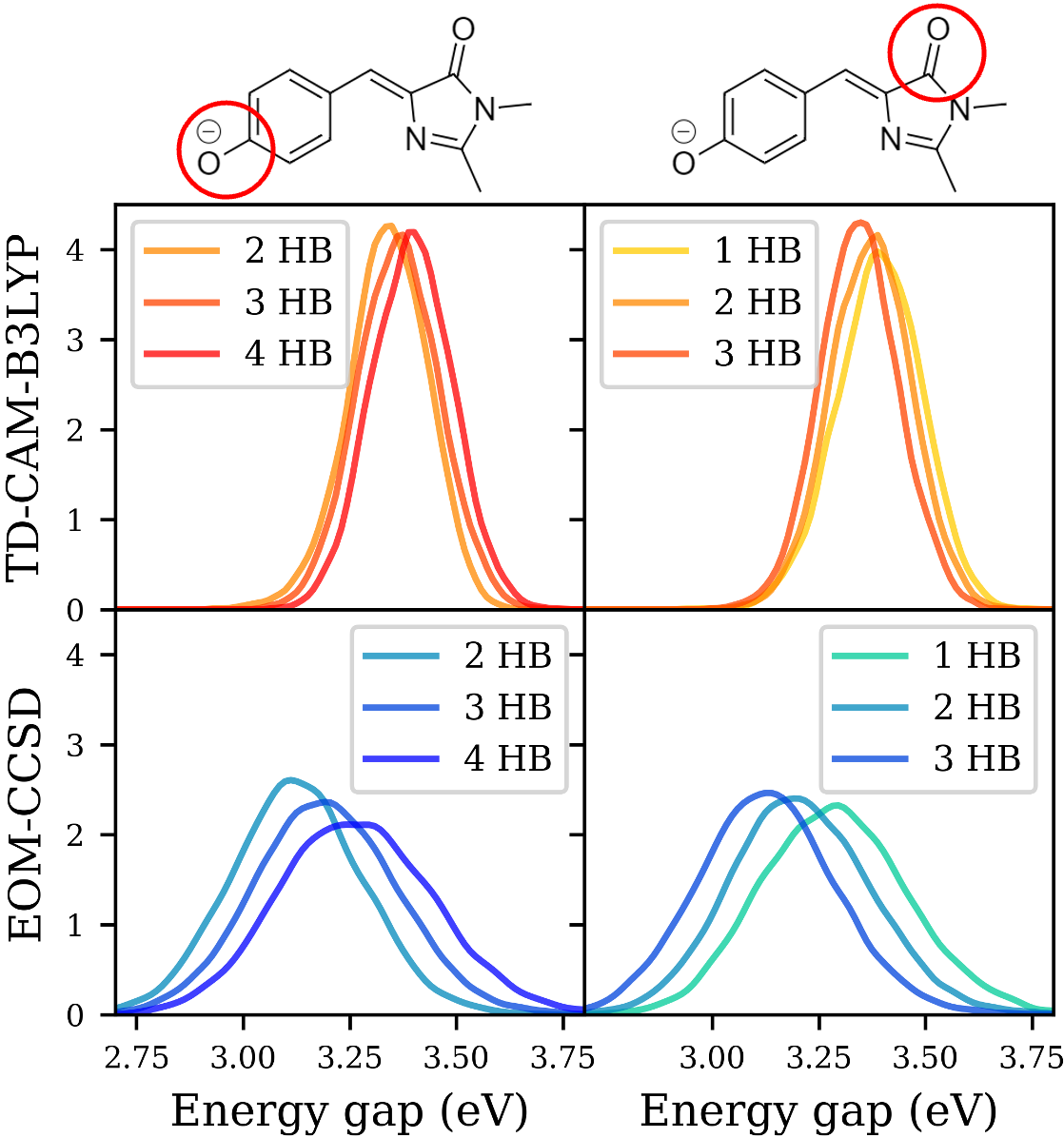}
    \end{center}
    \caption{Comparison of the hydrogen-bond dependent energy gap distributions for TD-CAM-B3LYP (top) and EOM-CCSD (bottom) energy gaps, as predicted by our ML models, conditioned on the number of hydrogen bonds formed with the anionic \ce{C-O-} (left) and \ce{C=O} (right) moieties of the GFP chromophore. The EOM-CCSD energy gap probability distributions are significantly more sensitive to the hydrogen bonding environment.}
	\label{fig:hb}
\end{figure}

To uncover the chemical origins of the solvent-induced broadening of the GFP chromophore in water upon using the higher level EOM-CCSD treatment of its electronic structure, we consider the effect of hydrogen bonding to the chromophore on the electronic excitation at the two primary hydrogen bonding sites on the GFP chromophore anion: the \ce{C-O-} group on the phenolate ring ((P)-ring) and the C=O group on the imidazole ring ((I)-ring). Figure~\ref{fig:hb} shows the distribution of electronic energy gaps between the ground and first excited state grouped by the number of hydrogen bonds formed to the primary hydrogen bonding sites on the GFP chromophore (see SI Sec.~\ref{si:sec:hb-criteria}). We note that although increasing the number of hydrogen bonds formed to the \ce{C-O-} increases the mean electronic excitation energy, the opposite shift in the excitation energy occurs as the number of hydrogen bonds to the C=O group in increased. Comparing the top (TD-CAM-B3LYP) and bottom (EOM-CCSD) panels of Figure~\ref{fig:hb}, we observe that TD-CAM-B3LYP predicts both a much smaller distribution of excitation energies for each hydrogen bonding environment as well as a much smaller shift in the mean of the excitation energy distribution upon changing the number of hydrogen bonds to the respective hydrogen bonding oxygen. In particular, the difference in the mean excitation energy upon going from two hydrogen bonds to the \ce{C-O-} group to four gives a change of only 0.06 eV with TD-CAM-B3LYP but for EOM-CCSD gives a 2.3 fold larger shift of 0.14 eV. A similar situation is observed for the C=O group where the difference in the mean excitation energy between forming 1 and 3 hydrogen bonds is -0.04 eV when using TD-CAM-B3LYP and -0.16 eV using EOM-CCSD, a 4-fold difference. The combination of hydrogen bonds that gives rise to the smallest mean excitation energy is obtained when three hydrogen bonds are formed to C=O and only two are formed to \ce{C-O-}. The largest mean excitation is observed in the opposing extreme case (see SI Sec.~\ref{si:sec:double-hb-envs}) where the number of hydrogen bonds formed to \ce{C-O-} is (4) and the number formed to C=O is (1). For TD-CAM-B3LYP the mean excitation energy for these two cases are 3.31 and 3.41~eV, a difference of 0.10 eV, whereas for EOM-CCSD they are 3.05 and 3.35 eV, a difference of 0.30 eV. Hence there is a three-fold greater shift in the mean electronic excitation energy when using EOM-CCSD upon changing the hydrogen bonding to the two primary hydrogen bonding sites on the chromophore. These results strongly indicate that the greater response of the electronic excitations on the GFP chromophore anion to the hydrogen bonding environment using EOM-CCSD plays a major role in understanding its broad optical absorption spectrum. 

The reason for the shifts observed in the electronic energy gap upon hydrogen bonding can be understood by considering the nature of the S$_0 \rightarrow$ S$_1$ electronic excitation that results in electron density being transferred from the (P)- to the (I)-ring. Increased hydrogen bonding to the \ce{C-O-} on the (P)-ring thus preferentially electrostatically stabilizes the ground state, which has a higher negative charge on the (P)-ring than the excited state, and thus elevates the excitation energy\cite{Shedge2019effect}. In contrast, forming more hydrogen bonds to the C=O on the (I)-ring stabilizes the extra electron density that ring receives upon the electronic excitation and thus provides increased electrostatic stabilization to the excited state relative to the ground state, in turn lowering the excitation energy. 

Both TD-CAM-B3LYP and EOM-CCSD exhibit relative stabilization of the ground and excited states by electrostatic interactions with the solvent that are heavily modulated by the hydrogen bonds formed between the GFP chromophore and water. In order to understand how the magnitude of this effect differs between the two electronic structure methods we thus consider the change in the difference dipole upon electronic excitation that facilitates this coupling. From Eq.~\ref{eq:mu_x} we can see that the difference dipole along a particular molecular axis connects the change of the energy gap, $\Delta U$, associated with the S$_0 \rightarrow$ S$_1$ excitation with the external field ($F_x$), arising from its solvation environment, experienced by the chromophore along that axis. The difference dipole distributions obtained from solution phase configurations of the chromophore thus serve as a measure of the sensitivity of the chromophore's electronic energy gap to changes in the external electric field along the defined axis, which in our case is the molecular frame axis defined by the center of the (P)- and (I)-rings on the GFP chromophore (see SI Sec.~\ref{si:sec:diff-dipole-calcs}).

\begin{align}
    \label{eq:mu_x}
    \Delta \mu_{x} \approx -\Delta U/ F_{x}
\end{align}

\begin{figure}[]
    \begin{center}
        \includegraphics[width=0.48\textwidth]{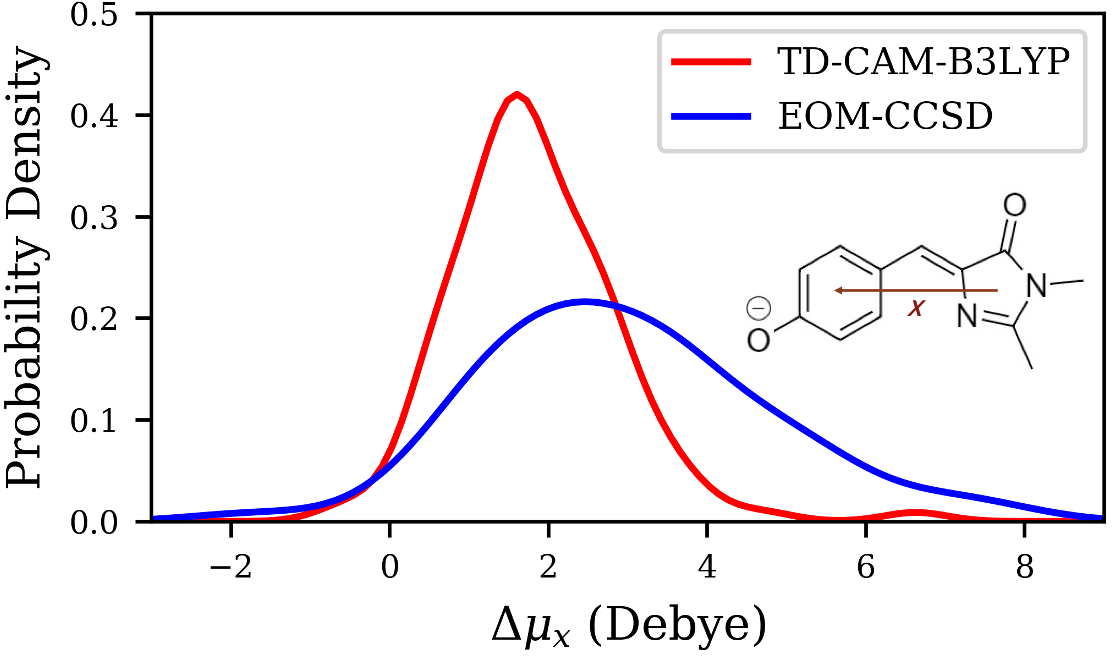}
    \end{center}
    \caption{Distributions of the excited-ground difference dipole ($\Delta \mu_{x}$) along the molecular frame axis shown in the inset for TD-CAM-B3LYP (red) and EOM-CCSD (blue) with the distribution of configurations of the GFP chromophore sampled from our aqueous phase simulations. The distribution for EOM-CCSD is broader and skewed toward higher $\Delta \mu_{x}$ compared to TD-CAM-B3LYP, which is indicative of a greater sensitivity, on average, of the chromophore's excitation with respect to changes in the external electrostatic environment arising from the solvent.}
	\label{fig:mux}
\end{figure}

Figure~\ref{fig:mux} compares the difference dipole, $\Delta \mu_{x}$, distributions for TD-CAM-B3LYP and EOM-CCSD. By comparing these distributions, we observe that EOM-CCSD on average possesses a larger difference dipole than TD-CAM-B3LYP (2.96 vs 1.83 D) and also gives a much broader distribution of difference dipoles. The larger average and range of difference dipoles for EOM-CCSD thus make its changes in the electronic energy gap,~$\Delta U$, more sensitive to the solvent electric fields, $F_x$, than TD-CAM-B3LYP, leading to a wider distribution of electronic energy gaps upon solvation and thus a broader electronic spectrum. Indeed, if we calculate the distribution of $\Delta U$, which for each configuration of the chromophore in solution is given as the difference of its electronic energy gaps computed in solution and in the absence of solvent, the ratio of the standard deviations of this for the two electronic structure methods, $\sigma_{\Delta U,\text{EOM-CCSD}}/\sigma_{\Delta U,\text{TD-CAM-B3LYP}}=1.761$, closely matches the ratio of the standard deviations for their difference dipole distributions, $\sigma_{\Delta \mu_x,\text{EOM-CCSD}}/\sigma_{\Delta \mu_x,\text{TD-CAM-B3LYP}}=1.757$ (see SI Sec.~\ref{si:sec:quantify_solv_broadening}). This result follows from Eq.~(\ref{eq:mu_x}) on the assumption that the field arising from the solvent, $F_{x}$, is similar in both cases, which one might expect since in both cases the solvent is treated at the DFT level of theory. The change in the difference dipole distribution is thus able to explain the larger hydrogen-bond dependent shift in the electronic energy gaps we see in the EOM-CCSD calculations and ultimately why compared with TD-CAM-B3LYP it gives a broader and more accurate linear electronic absorption spectrum for the GFP chromophore in water. Such an explanation is consistent with the even more extreme case of protonation of the gas-phase chromophore where the shift upon protonation is much higher for EOM-CCSD (1.03 eV) than for TDDFT (0.38 eV) \cite{Filippi_2009}.

In addition to our spectra obtained from classical ab initio molecular dynamics (AIMD) within the second order cumulant approach, we also investigated the effect of incorporating NQEs using \textit{ab initio} thermostatted ring polymer molecular dynamics (TRPMD)\cite{Craig2004quantum,Habershon2013ring,Rossi2014how} trajectories\cite{Zuehlsdorff2018unraveling}. From the middle panel of Fig.~\ref{fig:specdens_linspec} and SI Figure \ref{si:fig:rpmd}, we see that including NQEs leads to a significant red shift in the peak position of the linear absorption spectrum for the solvated GFP chromophore that is more pronounced for EOM-CCSD (3.16 $\rightarrow$ 3.07~eV) than for TDDFT (3.30 $\rightarrow$ 3.27~eV), bringing the simulated peak position closer to that of the experimental spectrum (2.94~eV). As shown in SI Figure~\ref{si:fig:hb-rpmd} the larger red shift for EOM-CCSD upon including NQEs arises from a shift in the hydrogen bond distribution with fewer hydrogen bonds being formed to the \ce{C-O-} group on the GFP chromophore, which corresponds to configurations that lead to lower excitation energies. 

\begin{figure}[]
    \begin{center}
        \includegraphics[width=0.5\textwidth]{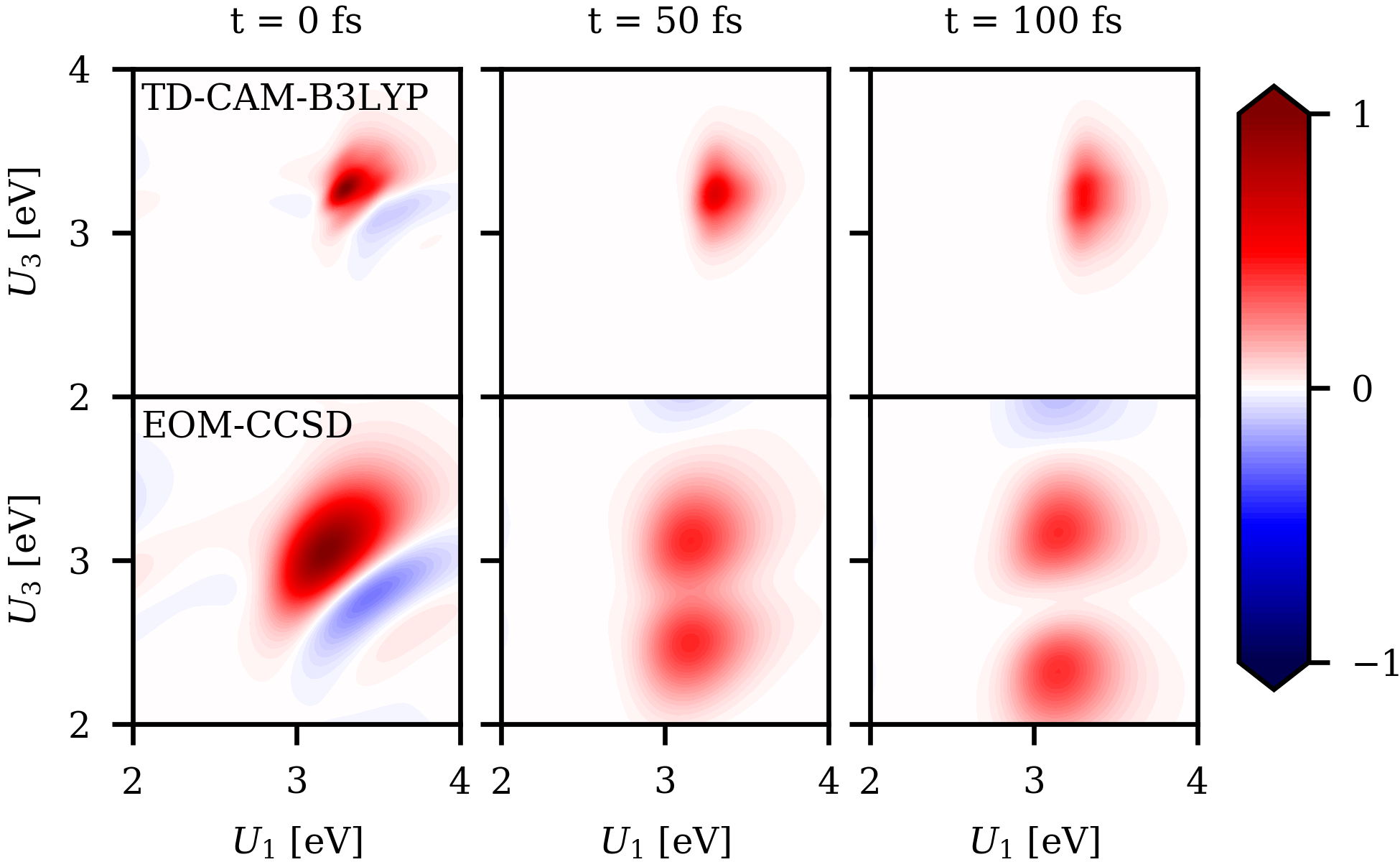}
    \end{center}
    \caption{Comparison of the two-dimensional electronic spectra (2DES) at three different time delays, t = 0, 50, 100~fs, using electronic energy gaps predicted by ML models trained to TD-CAM-B3LYP vs.~EOM-CCSD showing that the dynamic Stokes shift, i.e., the separation between the diagonal ground state bleach and the stimulated emission peak below the diagonal, is significantly more pronounced for the predicted EOM-CCSD spectra.}
	\label{fig:2des}
\end{figure}

Finally, given the success of the EOM-CCSD method in capturing the linear optical spectra of the GFP chromophore in aqueous solution, in Figure~\ref{fig:2des} we also compare the simulated 2D electronic absorption spectra (2DES)\cite{Mukamel2000multidimensional,Jonas2003two,Cho2008coherent} obtained using our ML models of TD-CAM-B3LYP and EOM-CCSD. The most striking difference between the electronic structure methods is the extent of the Stokes shift that emerges in the EOM-CCSD spectrum, where even by a delay time of 50~fs we already see that the ground state bleach (on the diagonal) and stimulated emission (below the diagonal) peaks have become well-separated. In contrast, the two peaks for TD-CAM-B3LYP still overlap considerably even at a delay time of 100~fs. It is important to note that previous computational studies of the GFP chromophore in water have found the chromophore undergoes twisted intramolecular charge transfer in the excited S$_1$ electronic state, with either the (P)- or (I)- rings often rotating in approximately 200-400~fs after the excitation\cite{Jones2021resolving}. In contrast, our cumulant-based 2DES are calculated using equilibrium molecular dynamics trajectories sampled on the electronic ground state (i.e. within the Franck-Condon region) and thus cannot account for the twisting of the chromophore in the excited state. Hence here we only report the short time 2DES up to 100~fs, which already highlights the differences between the two electronic structure methods.

In summary, we have introduced a data-efficient transfer learning protocol for training ML models to predict the electronic excited state energy gaps for chromophores in solution. This protocol uses data from low-level electronic structure calculations to train an initial set of ML models that are then used to initialize the training of models to the energy gaps obtained using a higher-level method. As a proof of concept, we employed this procedure to train ML models for energy gaps of the GFP chromophore in water at the level of EOM-CCSD embedded in a DFT description of the solvent. The training set for these models used only 400 embedded EOM-CCSD energy gaps, which is 10 times fewer than the training data needed to train models for similar systems in the past where we did not use transfer learning\cite{Chen2020exploiting}. We expect that this transfer learning protocol will afford similar data efficiency gains for chromophores in other complex environments, such as other solvents or protein scaffolds.

We showed that the linear spectrum obtained from our ML model trained to embedded EOM-CCSD energy gaps better reproduces the experimental spectrum whereas TD-CAM-B3LYP severely underpredicts the spectral width. The source of this difference arises from the inability of TD-CAM-B3LYP to capture the extent to which the energy gap shifts due to the hydrogen bonding environment. Underlying this difference is an inherently different description of the sensitivity of the chromophore's energy gap to environmental electric field effects, which also leads to a more pronounced red shift when NQEs are included and brings the optical absorption spectrum into even better agreement with experiment. Hence, for the GFP chromophore in water, and other systems where the chromophore is involved in strong hydrogen bonding interactions with its environment, it is important to employ higher-level electronic structure methods and NQEs in order to accurately simulate optical absorption spectroscopies.

\section*{Acknowledgments}
This work was funded by the U.S. Department of Energy, Office of Science, Office of Basic Energy Sciences (DE-SC0020203 to C.M.I. and T.E.M.). Y.M. thanks Prof.~Anna Krylov (USC) for helpful suggestions on EOM-CCSD calculations. This research used resources of the National Energy Research Scientific Computing Center (NERSC), a U.S. Department of Energy Office of Science User Facility located at Lawrence Berkeley National Laboratory, operated under Contract No. DE-AC02-05CH11231 using NERSC Awards BES-ERCAP0019987 and BES-ERCAP0023755.

\bibliography{bibliography}

\end{document}


\title{Supplementary material for ``Elucidating the role of hydrogen bonding in the optical spectroscopy of the solvated green fluorescent protein chromophore: using machine learning to establish the importance of high-level electronic structure''}

\author{Michael S. Chen}
\affiliation{Department of Chemistry, Stanford University, Stanford, California 94305, USA}

\author{Yuezhi Mao}
\affiliation{Department of Chemistry, Stanford University, Stanford, California 94305, USA}

\author{Andrew Snider}
\affiliation{Chemistry and Biochemistry, University of California Merced, Merced, California 95343, USA}

\author{Prachi Gupta}
\affiliation{Chemistry and Biochemistry, University of California Merced, Merced, California 95343, USA}

\author{Andr\'es Montoya-Castillo}
\affiliation{Department of Chemistry, University of Colorado, Boulder, Boulder, Colorado 80309, USA}

\author{Tim J. Zuehlsdorff}
 \affiliation{Department of Chemistry, Oregon State University, Corvallis, Oregon 97331, USA}

\author{Christine M. Isborn}
\email{cisborn@ucmerced.edu}
\affiliation{Chemistry and Biochemistry, University of California Merced, Merced, California 95343, USA}

\author{Thomas E. Markland}
\email{tmarkland@stanford.edu}
\affiliation{Department of Chemistry, Stanford University, Stanford, California 94305, USA}

\date{\today}

\maketitle

\tableofcontents

\onecolumngrid
\normalsize

\section{Electronic structure}
\label{si:sec:electronic_structure}
\subsection{Previous studies of the excited states of the HBDI$^-$ chromophore }

The HBDI$^-$ chromophore has been extensively studied computationally in vacuum,\cite{Martin2004, Altoe_2005, Epifanovsky_2009, Filippi_2009, Davari_2014, Georgieva2017high, List2022} solvated, \cite{Toniolo2004, Altoe_2005, Polyakov_2009, Avila_Ferrer_2014, Zutterman_2017, Georgieva2017high, Zuehlsdorff2018combining, Zuehlsdorff2018unraveling, Raucci_2020, Zutterman_2022} and native protein environments.\cite{Voityuk_1997, Filippi2012, Grigorenko_2013, Acharya_2017, Jones_2022} In gas phase, the anion has an absorption maximum around 2.56 eV, \cite{Nielsen2001, Andersen2004} whereas the neutral form has a maximum near 3.65 eV, giving an extremely large shift of well over 1 eV.\cite{Greenwood2014} Previous work studying both the anionic and neutral form of the chromophore in vacuum shows that TDDFT significantly underestimates the difference in excitation energy between the anionic and neutral species, whereas EOM-CCSD, complete-active-space second-order perturbation theory (CASPT2), and quantum Monte Carlo (QMC) methods all correct this shortcoming, with CASPT2 predicting a shift of 0.76 eV and EOM-CCSD predicting a shift of 1.03 eV.\cite{Filippi_2009} In gas phase, the simulated vibronic spectral shapes of the anionic chromophore are very similar for both high-level wave function methods and for TDDFT methods that have sufficient exact exchange, although calculations of the spectrum employing CASSCF showed a larger vibronic progression than other methods, which the authors claim is an artifact due to the neglect of dynamic electron-electron correlation.\cite{Davari_2014} High level calculations using resolution-of-the-identity algebraic diagrammatic construction through second-order (RI-ADC(2)) of both the anionic and neutral forms of the chromophore with microsolvation showed that the electronic structure of the anionic chromophore becomes more similar to that of the neutral form upon being solvated, leading to the blue shift in the absorption spectrum.\cite{Georgieva2017high} 

The spectrum of the chromophore in aqueous solution is much broader than in the protein, and this broadening has been underestimated in many studies,\cite{Avila_Ferrer_2014, Zuehlsdorff2018combining, Zuehlsdorff2018unraveling, Raucci_2020} most of which used time-dependent density functional theory to model the excitation energy of the chromophore in water. In work by Avila Ferrer et al. modeling the vibronic spectrum, the authors hypothesized that the missing broadening was due to population of non-planar geometries.\cite{Avila_Ferrer_2014} A study by Raucci et al. included full sampling of chromophore-solvent interactions that included non-planar geometries, and they attributed the missing broadening to the lack of vibronic effects.\cite{Raucci_2020} However, in studies by some of the current authors\cite{Zuehlsdorff2018combining, Zuehlsdorff2018unraveling} that included both vibronic effects and full sampling of chromophore-solvent non-planar geometries, as well as nuclear quantum effects and a QM treatment of the explicit solvent environment in the QM region of the TDDFT calculation, the UV-Vis absorption lineshape was improved, but still underestimated in spectral width, suggesting an alternative explanation for the missing broadening.

\subsection{TDDFT and CIS computational details}

In this work, excitation energies were computed for the HBDI$^-$ chromophore using time-dependent density functional theory (TDDFT) \cite{Casida1995,Dreuw2005single} within the Tamm-Dancoff approximation (TDA) \cite{Hirata1999time} with the range-separated hybrid CAM-B3LYP \cite{Yanai2004new} functional and 6-31+G(d) \cite{Frisch1984self}  basis set, as well as with the configuration interaction singles (CIS) \cite{Bene1971self,Foresman1992toward} method with the same basis set. For the solvated chromophore, 167 water molecules were included in the QM region. The QM region was surrounded by implicit solvent using the COSMO \cite{Klamt1993cosmo} solvation model with the water dielectric constant and atomic radii scaled by a value of 1.5 for the solute cavity. The GPU-accelerated TeraChem electronic structure program\cite{Seritan_TeraChem_2021} was used for all TDDFT and CIS calculations with integral threshold cutoffs of 1e-13.  

\subsection{EOM-CCSD and projection-based embedding}

Equation-of-motion coupled-cluster with singles and doubles (EOM-CCSD) \cite{Stanton1993equation, Comeau1993equation, Krylov2008equation} is a correlated wavefunction method for electronic excited states. It has been shown to afford excellent accuracy for many types of electronic excitations, \cite{Loos2018mountaineering,Loos2020mountaineering} and thus here we employ the energy gaps predicted using EOM-CCSD as the final target of our ML models. Due to the steeply increasing computational cost of EOM-CCSD with system size (scales as $\mathcal{O}(N^6)$), it is unfeasible to treat the entire chromophore-solvent system using this method. Instead, we employ an EOM-CCSD excited-state calculation on the HBDI$^-$ chromophore and treat the surrounding solvent molecules at the level of ground-state density functional theory (DFT). This was achieved by employing our implementation of the projection-based embedding theory originally developed by Manby, Miller, and co-workers \cite{Manby2012simple,Lee2019projection} in the Q-Chem 5 software package,\cite{Epifanovsky2021} which allows wavefunction theory (WFT) calculations to be performed in environments described using DFT. Note that EOM-CCSD-in-DFT embedding has been previously employed by others \cite{Bennie2017pushing,Wen2019absolutely,Parravicini2021embedded} to capture electronic excitations in solvent or other condensed-phase (e.g., protein) environments. With our Q-Chem implementation, an embedded EOM-CCSD calculation for a solute-solvent system comprises the following steps:
\begin{enumerate}
    \item Perform a standard ground-state DFT calculation for the full system to obtain the canonical occupied and virtual orbitals.
    \item Localize the canonical occupied orbitals and select the ones belonging to the chromophore that will remain active in the embedding calculation. In this work, we employ the recently proposed subsystem-projected atomic orbital decomposition (SPADE) \cite{Claudino2019automatic} approach to achieve this partition. The occupied orbitals assigned to the environment will no longer be changed in the following steps.
    \item Perform an embedded Hartree-Fock (HF) for the chromophore (note that all virtual orbitals of the full system are relaxed during this step). Once converged, to reduce the cost of the subsequent WFT calculation, we further truncate the space spanned by the resulting virtual orbitals using the concentric localization (CL) \cite{Claudino2019simple} scheme, a systematic way to tune the accuracy and cost of embedded WFT calculations by including more/less virtual orbitals.
    \item With both the environment occupied and excluded virtual orbitals frozen, perform an embedded EOM-CCSD calculation based on the HF reference to obtain the excitation energy of the chromophore in the solvation environment.
\end{enumerate}

\begin{figure*}[h]
    \begin{center}
        \includegraphics[width=0.5\textwidth]{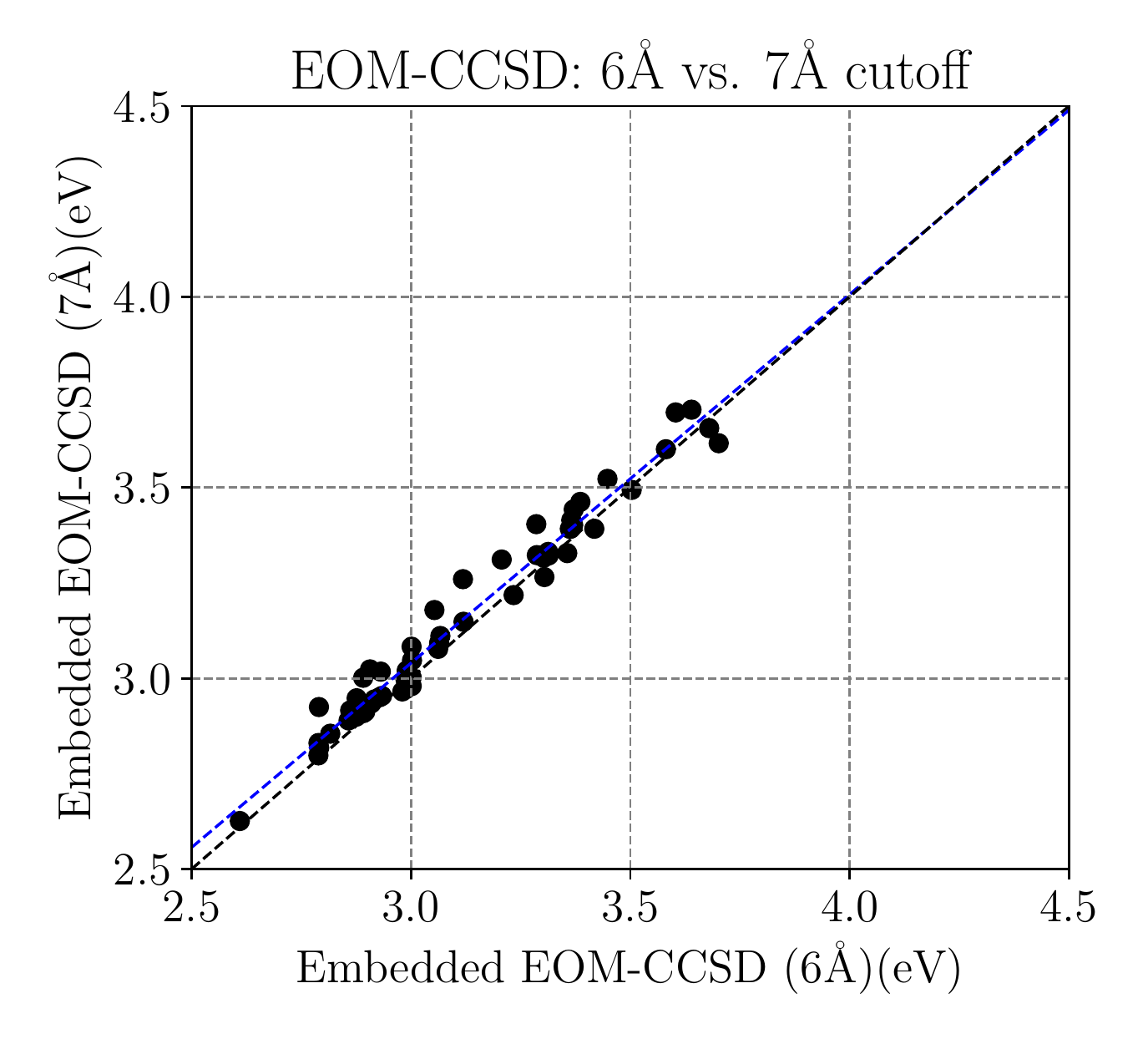}
    \end{center}
    \caption{Correlation plot comparing embedded EOM-CCSD excitation energies for the solvated GFP chromophore when using a 6 vs. 7 \AA\ cutoff radius to determine which solvent water molecules to include in the calculations. The smaller cutoff only introduces small changes to the excitation energies.}
	\label{si:fig:eEOMCCSD-cutoffcheck}
\end{figure*}

We performed the embedded EOM-CCSD calculations on truncated HBDI$^-$--water clusters, in which the oxygens of the solvent water molecules are within a range of 6 \AA\ from at least one of the solute atoms. To examine whether this relatively small cutoff radius (6 \AA) has a negative impact on the accuracy of these embedding calculations, we repeated the calculations on a subset of the training configurations (60 frames) but with a 7 \AA\ cutoff. As shown by excellent correlation in Figure~\ref{si:fig:eEOMCCSD-cutoffcheck}, using a larger cutoff radius introduces insignificant changes to the resulting embedded EOM-CCSD excitation energies. The embedding calculations were carried out using the 6-31+G(d) \cite{Frisch1984self} basis set, and the PBE0 \cite{Adamo1999toward} functional was chosen to describe the environment (i.e., the ``low-level'' theory in the embedding calculation). For the truncation of the virtual space using the CL scheme, the projection basis (PB, as defined in Ref.~\citenum{Claudino2019simple}) was chosen to be the same as the working basis (WB, 6-31+G(d)), and only the first CL shell was included due to the formidable computational cost such that the number of active virtuals is equal to that of the WB functions on the HBDI$^-$ chromophore (which are both 326). In addition, the EOM-CCSD calculations were conducted with the frozen-core approximation, i.e., the $1s$ electrons of the C, N, and O atoms on the GFP chromophore are presumed not to participate its S$_0 \rightarrow$ S$_1$ excitation. These core orbitals remain inactive together with the environment occupied orbitals in the EOM-CCSD calculation. Overall, our procedure yields a (41o, 326v) active space for the embedded EOM-CCSD calculations for the HBDI$^-$--water clusters.

\begin{figure*}[h]
    \begin{center}
        \includegraphics[width=0.95\textwidth]{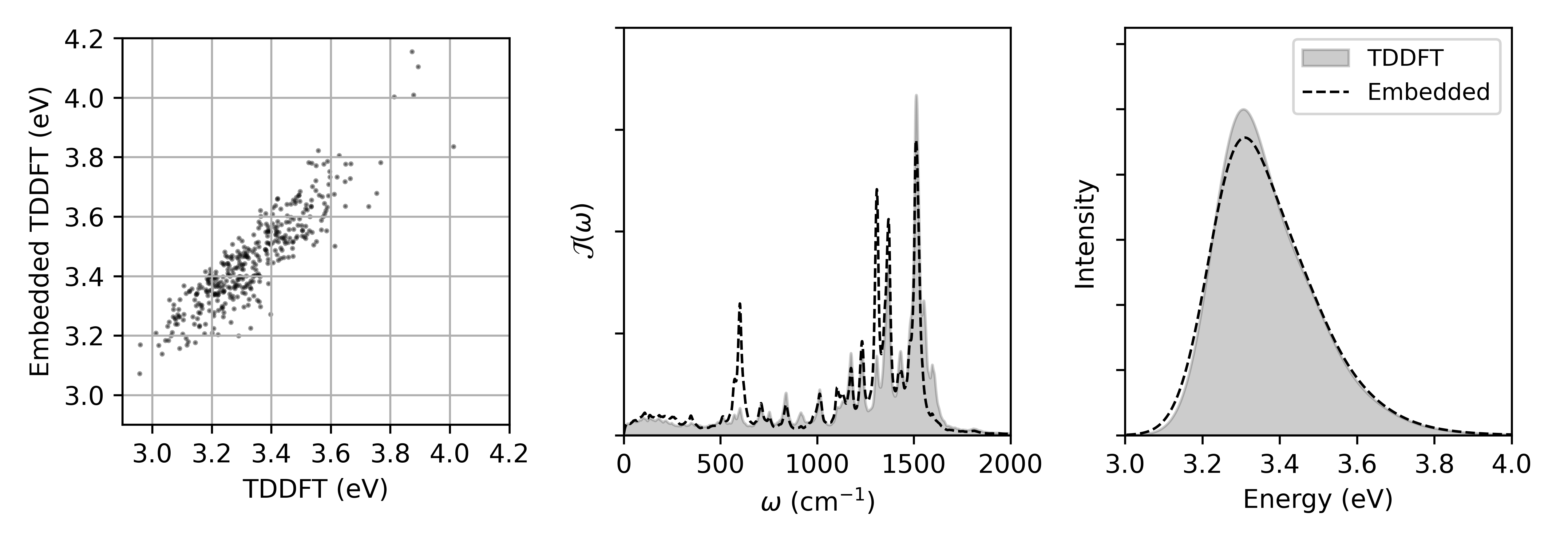}
    \end{center}
    \caption{Checks showing that any artifacts introduced by the embedding procedure do not significantly effect the shape of the linear spectrum. (left) Correlation plot of energy gaps for the 374 configurations computed using TDDFT and TDDFT embedded in DFT showing a strong correlation between the two (Pearson correlation coefficient of 0.89) and a systematic blue shift of energy gap values introduced by the embedding procedure. The spectral density (middle) and linear spectrum (right) that we obtain from a transfer learned ML model trained to 300 embedded TDDFT energy gaps, starting from a gradient-fitted CIS model, are in good agreement with those that we obtained from a gradient-fitted ML model trained to full TDDFT. We shifted the linear spectrum for embedded TDDFT so that the peak position aligns with that of TDDFT in order to compare the shapes of the two spectra.}
	\label{si:fig:eTDDFT-check}
\end{figure*}

To investigate whether the setup of our embedding calculations is associated with artifacts that can give rise to the substantial difference between the machine-learned TDDFT and EOM-CCSD spectra, we further enabled embedded TDDFT-in-DFT calculations in Q-Chem, in which the single excitation amplitudes are restricted to be in the same (41o, 326v) active space such that the excitations are centered on the chromophore. The procedure of these calculations largely follows the steps as described above, except that (i) an embedded CAM-B3LYP \cite{Yanai2004new} (the functional we use for standard TDDFT calculations) calculation is performed in Step 3 to generate the reference orbitals instead of HF, and (ii) the frozen-core approximation is not invoked in these embedded TDDFT calculations. Note that the Tamm-Dancoff approximation was also employed in the embedded TDDFT calculations as in the TD-CAM-B3LYP calculations performed with TeraChem.  Figure.~\ref{si:fig:eTDDFT-check} shows that although the energy gaps we obtain from our embedded TDDFT calculations are slightly blue-shifted, the spectral density and linear spectrum we obtain from a transfer learning model trained to embedded TDDFT energy gaps, following the same protocol as we have used for training our EOM-CCSD models (SI Sec.~\ref{si:sec:transfer-learning}), are in good agreement with what we have obtained using our accurate gradient-trained TDDFT ML model. From these results, we see our transfer learning model trained to embedded TDDFT gives us a linear spectrum whose shape is faithful to what we obtain when treating the full system with TDDFT, and that the embedding procedure is not introducing artifacts that manifest significantly in the linear spectrum.

\subsection{Difference electronic dipole calculations}
\label{si:sec:diff-dipole-calcs}

To shed light on the the difference in the sensitivity of TD-CAM-B3LYP and EOM-CCSD energy gaps to the hydrogen-bond environment, we calculated the difference dipole between the S$_1$ excited state and the S$_0$ ground state of the GFP chromophore (HBDI$^-$): \begin{equation}
    \Delta \boldsymbol{\mu} = \boldsymbol{\mu}_{\text{S}_1} - \boldsymbol{\mu}_{\text{S}_0},
\end{equation}
which can be used to quantify the change in chromophore's electronic energy gap o induced by per unit change in the solvent electric field and is also known as the electronic Stark tuning rate.

\begin{figure*}[h!]
    \begin{center}
        \includegraphics[width=0.8\textwidth]{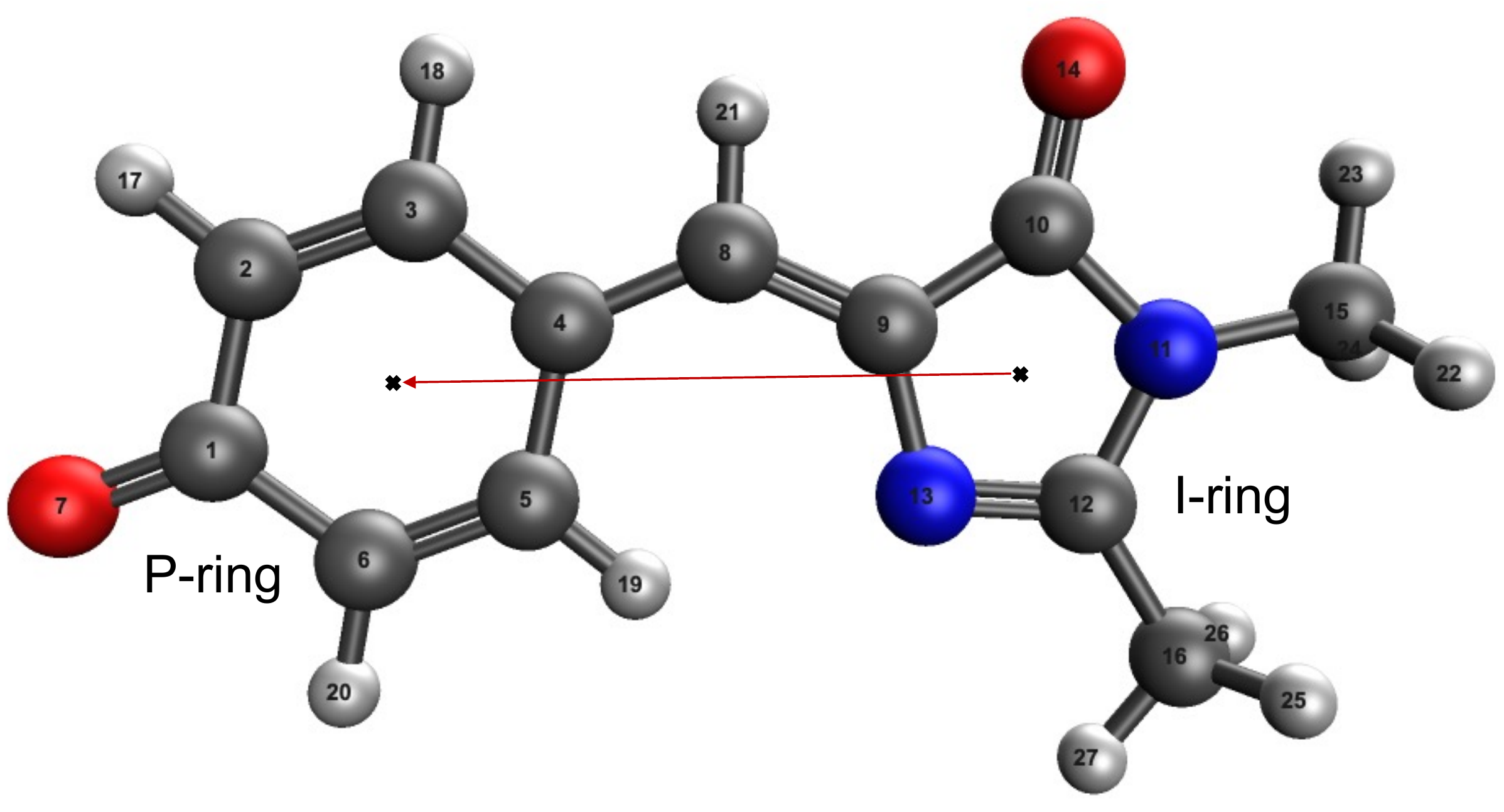}
    \end{center}
    \caption{Illustration of the ``charge-transfer axis'' that is used to generate the projected difference dipole ($\Delta \mu_x$) for each configuration.}
	\label{si:fig:ct_axis}
\end{figure*}

All electronic structure calculations involved in the evaluation of the difference dipole were performed using the Q-Chem 5.4 software package \cite{Epifanovsky2021} with the 6-31+G(d) basis set. For TD-CAM-B3LYP, $\boldsymbol{\mu}_{\text{S}_0}$ and  $\boldsymbol{\mu}_{\text{S}_1}$ were derived from the ground-state electron density and the \emph{relaxed} electron density of the TDDFT/TDA S$_1$ state, respectively, and for EOM-CCSD they were derived from the \emph{relaxed} electron densities of the CCSD ground state and the EOM-CCSD S$_1$ state (with keyword \texttt{CC\_FULLRESPONSE = 1}). We calculated the difference dipoles for 113 (randomly selected) out of the 375 configurations for GFP chromophore in water, in which the water molecules are stripped away while the GFP chromophore is fixed at its structure as in the MD snapshot. The $\Delta \boldsymbol{\mu}$ obtained for each configuration is then projected onto the ``charge-transfer axis'' (denoted as the $x$-axis), which in our study is defined by the vector pointing from the center of the I-ring (the average position of C9, C10, N11, C12, and N13) to that of the P-ring (the average of the coordinates of C1--C6) as illustrated in Fig.~\ref{si:fig:ct_axis}, and in Fig.~\ref{fig:mux} of the main paper we show the smoothed distribution of $\Delta \mu_x$ over the 113 configurations generated using the Gaussian kernel density estimation (KDE) functionality available in SciPy.

\section{Machine learning details}
\subsection{Descriptors and architectures}
Our ML models in this work use Behler-Parrinello neural networks\cite{Behler2007generalized,Behler2011atom,Behler2015constructing}, which are commonly used for the fitting of ground state potential energy surfaces. Here we train these models to instead predict the S$_0 \rightarrow$ S$_1$ energy gap of the GFP chromophore. In our ML models, the predicted energy gap of a given configuration, $U_{ML}$, is given as a sum of atomic contributions from the $N$ atoms in the system,
\begin{align}
    U_{ML} = \sum_{i=1}^N U_i = & \sum_{i=1}^N f_{NN}^{(k_i)} \left( \bm{G}_{i} ( \{ \bm{x} \} ) \right),
    \label{si:eq:energy-gap-sum}
\end{align}
where each of the atomic contributions is predicted by a neural network $f_{NN}^{(k_i)}$ specific to the atom type $k_i$ of atom $i$. The input to $f_{NN}^{(k_i)}$ is a set of descriptors $G_i$ that represents the local chemical environment around atom $i$ and is a function of the set of atomic positions, $\{ \bm{x} \}$, in a configuration. Here, we employ a modified version of the Chebyshev polynomial descriptors\cite{Artrith2017efficient} as implemented in the aenet software package\cite{Artrith2016implementation}. The Chebyshev polynomial descriptors implemented in aenet consist of radial descriptors for each atom $i$ that encode local pairwise interactions,
\begin{align}
    c_{i,\alpha}^{(2)} = \sum_{j}\phi_{\alpha}(R_{ij})f_c(R_{ij})w_j,
\end{align}
and angular descriptors that encode local 3-body interactions,
\begin{align}
    c_{i,\alpha}^{(3)} = \sum_{j,k}\phi_{\alpha}(\cos{\theta_{ijk}})f_c(R_{ij})f_c(R_{ik})w_jw_k,
\end{align}
where $R_{ij}$ is the distance between atoms $i$ and $j$, $\theta_{ijk}$ is the angle that atoms $j$ and $k$ make with atom $i$, $\phi_{\alpha}$ is a Chebyshev polynomial of the first kind, with an expansion order of $\alpha$, and $f_c$ serves as a smoothly decaying cutoff function of the form with a specified cutoff of $R_c$,

\begin{align}
    f_c = \begin{cases}
        0, &\text{ if $R_{ij} \geq R_c$}\\
        \frac{1}{2}\left[\cos(\pi R_{ij}/R_c) + 1\right], &\text{ if $R_{ij} < R_c$}.
    \end{cases}
    \label{eqn:cutoff-func}
\end{align}

Note that the atom types of atoms $j$ and $k$ are encoded as integer weights $w_j$ and $w_k$. aenet's Chebyshev polynomial descriptors by default consist of a concatenation of structural descriptors, where all weights $w_j$ and $w_k$ are 1, and type-specific descriptors with $w_j$ and $w_k$ taking on integer values specific to their atom type. The integer assignments are made arbitrarily, with atom types assigned to integers 0, $\pm1$, $\pm2$, where 0 is omitted when the system has an even number of different atom types.

In constructing our ML models for GFP in water, we specified a set of 6 different atom types $\{$H, C, N, O, O$_w$, H$_w$ $\}$ where the first four are specific to atoms on the GFP chromophore and the last two represent the oxygen and hydrogen atoms on the water molecules. Instead of using aenet's weight assignments, for the structural descriptors we set weights for all chromophore atoms to 1 and for all water atoms to 0, and for the type-specific descriptors we use $\{0.0, 1.0, 2.0, 3.0, 0.25, 0.50\}$, respectively. For all atom types, we use a max Chebyshev polynomial expansion order $\alpha=15$ for the radial descriptors and $\alpha=5$ for the angular descriptors as well as cutoffs of $R_c=6.00$~\AA\ and $R_c=3.00$~\AA, respectively. Our number of descriptors for a given atom, and consequently the number inputs to our neural networks, thus totals 44 because we have 22 for structural descriptors and 22 for type-specific descriptors (each consisting of 16 radial and 6 angular descriptors).

Each atom type specific neural network $f_{NN}^{(k_i)}$ we used was composed of two hidden layers with 25 hidden nodes per layer with a total 1,825 parameters. With 6 different atom types, each ML model thus consists of 10,950 parameters. We employed a committee of 8 ML models, each with the same architecture but with different weight initializations, using the committee's mean prediction for calculating spectra. 

\subsection{Datasets}
In order to train our ML models and benchmark procedures, we employed various datasets for both the gas-phase and solvated GFP chromophore at different levels of electronic structure theory. 

The 100 configurations comprising our gas-phase dataset were drawn from previously simulated AI-PIMD trajectories\cite{Zuehlsdorff2018unraveling} with a 800~fs spacing between selected configurations. The S$_0$ and S$_1$ energies as well as the corresponding gradients were computed using CIS, TD-CAM-B3LYP, and EOM-CCSD (SI Sec.~\ref{si:sec:electronic_structure}). The CIS dataset consisted of only 99 configurations since one of the calculations did not converge properly.

Our initial dataset for the GFP chromophore in water consisted of 375 configurations that were also drawn from previously simulated AI-PIMD trajectories\cite{Zuehlsdorff2018unraveling}, with a 150~fs spacing between selected configurations. The S$_0$ and S$_1$ energies as well as the corresponding gradients were computed using CIS and TD-CAM-B3LYP, whereas the embedded EOM-CCSD dataset only consisted of S$_0$-S$_1$ energy gaps (see SI Sec.~\ref{si:sec:electronic_structure}). The CIS dataset consisted of only 372 configurations since 3 of the calculations did not converge properly. The initial embedded EOM-CCSD dataset consisted of only 235 configurations. Two query-by-committee\cite{Seungh1992query,Krogh1994neural} iterations were subsequently conducted, with each iteration selecting 100 new configurations from the same set of AI-PIMD trajectories to add to the dataset. These additional configurations were selected such that no two configurations were separated by less than 100~fs. The final embedded EOM-CCSD dataset thus consisted of 435 configurations and their corresponding energy gaps.

\subsection{Training procedure and benchmarks}
In order to mitigate overfitting when training our models, we monitored the energy gap error evaluated over a validation set at every training epoch and employed early-stopping. When training all of our gas-phase ML models, the validation set consisted of 25 configurations. When training our TD-CAM-B3LYP and CIS models for the GFP chromophore in water the validation set consisted of 75 configurations, whereas the validation set for our embedded EOM-CCSD models consisted of 35 configurations.

\subsubsection{Training on gradient data}
\label{si:sec:grad-training}
We trained ML models on energy gap gradients, i.e., excited minus ground state energy gradients with respect to nuclear positions, for the gas-phase and solvated GFP chromophores in order to benchmark and initialize our transfer learning procedure. These models are trained only on gradient data; therefore, in order to pin the ML surface we apply a constant shift to the predicted energy gap mean so that it matches the actual energy gap mean of the training set. For our gas-phase models, our loss function for the set of NN parameters $\theta$ and a single training configuration is the mean square deviation of the predicted gradients $\partial U_{ML}/\partial r_{i,k}$ with respect to the target gradients $\partial U/\partial r_{i,k}$,
\begin{align}
    L(\mathbf{\theta}) = \frac{1}{3N}\sum_{i=1}^{N}\sum_{k=x,y,z}\left( \frac{\partial U_{ML}}{\partial r_{i,k}} - \frac{\partial U}{\partial r_{i,k}} \right)^2.
\end{align}

\begin{figure*}[]
    \begin{center}
        \includegraphics[width=0.75\textwidth]{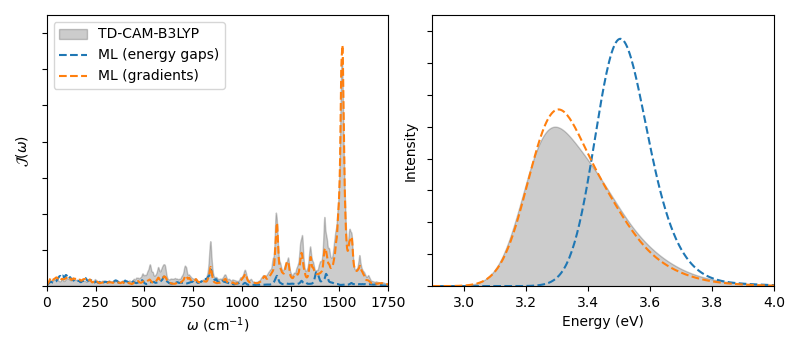}
    \end{center}
    \caption{For the GFP chromophore in water, our ML model trained on TD-CAM-B3LYP energy gap gradients (blue) from 300 configurations accurately reproduces the features of the reference TD-CAM-B3LYP spectral density (left) and the shape of the linear spectrum (right) shaded in gray. The ML model trained on only the energy gaps gives a spectral density where vibronic peak intensities are severely underpredicted, resulting in a linear spectrum that is missing its vibronic shoulder.}
	\label{si:fig:gas-phase-gradfit-check}
\end{figure*}

In SI Fig.~\ref{si:fig:gas-phase-gradfit-check} we benchmark the performance of our ML model trained to gradient data for the gas-phase GFP chromophore as compared to an ML model where only the energy gaps were used for training. The training set for both models consisted of 75 configurations, with the gradient-trained model achieving a much lower energy gap prediction root mean square error (RMSE) of 0.0276~eV for the 25 configuration validation set than the model trained only on energy gaps (0.0723~eV). Both the spectral density and the linear spectrum obtained from our ML model trained to gradients (blue line) show excellent agreement with the reference TD-CAM-B3LYP results in gray computed for a 14~ps long AIMD trajectory consisting of 7000 configurations.

For the solvated system we added a regularization term to the loss function in order to reduce solvent atom contributions to the energy gap,
\begin{align}
    L(\mathbf{\theta}) = \frac{1}{3N}\sum_{i=1}^{N}\sum_{k=x,y,z}\left( \frac{\partial U_{ML}}{\partial r_{j,k}} - \frac{\partial U}{\partial r_{j,k}} \right)^2 + \frac{\lambda}{N-M}\sum_{j=1}^{N-M} \mathcal{U}_j^2 ,
\end{align}
where $N-M$ is the number of solvent atoms, $\mathcal{U}_j$ is the predicted energy gap contribution for the $j^{th}$ solvent atom, and $\lambda$ is a user-specified weight for the regularization term. Without this regularization term, the resulting ML model for the energy gap is likely to make unphysically extensive predictions with respect to the number solvent atoms in a configuration. In practice, we employed a regularization weighting of $\lambda=10$ for our gradient-trained solvated ML models.

All gradient-trained ML models were initialized randomly using the Nguyen-Widrow initialization method\cite{Nguyen1990improving}. Parameters were optimized using the Adam optimizer\cite{Kingma2015Adam} with a learning rate of 0.001 and parameters were updated on a per configuration basis.

\begin{figure*}[]
    \begin{center}
        \includegraphics[width=0.5\textwidth]{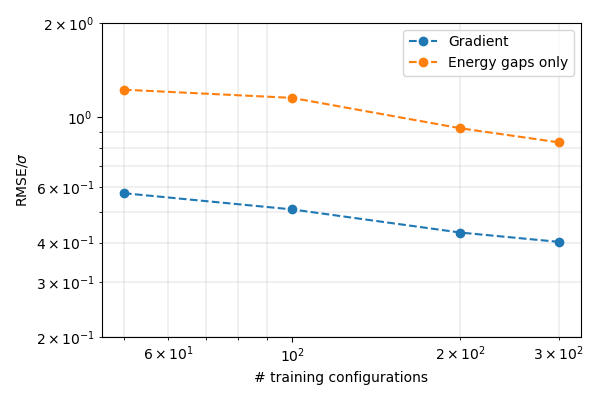}
    \end{center}
    \caption{A comparison of learning curves for our ML models trained on TD-CAM-B3LYP energy gaps (orange) versus energy gap gradients (blue) for the GFP chromophore in water showing how the energy gap prediction errors scale with training set size. Errors are reported as RMSEs over a validation set of 75 configurations that have been scaled by the standard deviation of energy gaps in our 375 configuration dataset ($\sigma=$0.165 eV).}
	\label{si:fig:solv-gradfit-check}
\end{figure*}

SI Fig.~\ref{si:fig:solv-gradfit-check} is a learning curve showing the accuracy gains we achieve when training our ML model on TD-CAM-B3LYP gradient data for the solvated system (i.e., GFP chromophore in water). For all training set sizes (50, 100, 200, and 300 configurations), we see that the gradient-trained models outperforms the models trained on only energy gaps with considerably lower RMSEs over our validation set, which consisted of the same 75 configurations for all training set sizes we scanned. To contextualize our validation set errors with respect to our previously published benchmark results, here we report the RMSEs as normalized by the standard deviation ($\sigma=0.165$ eV) for the energy gaps in our dataset. We previously showed that for a similar system, the PYP chromophore in water, that we could obtain accurate spectral densities and spectra using a model trained to only 2000 energy gaps that gave a validation set RMSE of 0.0505~eV (RMSE/$\sigma$ = 0.410)\cite{Chen2020exploiting}. Here with our model trained on gradient data from 300 configurations we achieve a similar normalized error of RMSE/$\sigma$ = 0.402.

\subsubsection{Transfer learning}
\label{si:sec:transfer-learning}
Our transfer learning procedure makes use of the hidden-solvent and indirect-solvent approaches that we previously introduced\cite{Chen2020exploiting}. In short, both ML models only sum over predicted chromophore atom contributions to the energy gap, i.e., $N$ in Eq.~\ref{si:eq:energy-gap-sum} would be the number of chromophore atoms in the system. However, the hidden-solvent model completely neglects solvent atom positions in the calculation of each chromophore atom $i$'s descriptor $\bm{G}_{i} ( \{ \bm{x} \} )$ while the indirect solvent model does consider solvent atom positions for the descriptor calculations.

\begin{figure*}[]
    \begin{center}
        \includegraphics[width=0.5\textwidth]{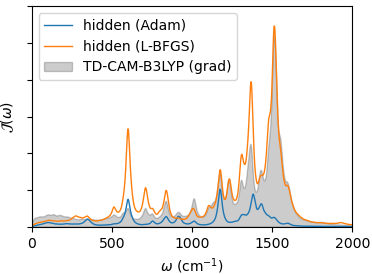}
    \end{center}
    \caption{A comparison of spectral densities for the GFP chromophore in water obtained from hidden-solvent ML models trained using the Adam (blue) or L-BFGS (orange) optimizers. Both models are trained using the same 300 TD-CAM-B3LYP energy gaps yet the model trained using the L-BFGS optimizer better captures the reference TD-CAM-B3LYP spectral density obtained via a gradient-trained model (gray), especially the peak intensities at higher frequencies.}
	\label{si:fig:lbfgs-vs-adam}
\end{figure*}

As diagrammed in Fig.~\ref{fig:summary_schematic}, the transfer learning models are initialized with weights taken from models trained to the energy gap gradients at a lower level of electronic structure theory (e.g. CIS or TD-CAM-B3LYP) from 300 configurations. The first transfer learning step is to train on the higher level energy gaps (e.g. embedded EOM-CCSD in our case) via a hidden-solvent approach. This training is conducted using the L-BFGS optimizer\cite{Liu1989limited} with a learning rate of 0.1 and using the strong Wolfe conditions for approximate line search\cite{Wolfe1969convergence,Wolfe1971convergence}. Parameter updates are made once per epoch in a fully-batched fashion. We found that using the L-BFGS optimizer for the hidden-solvent fits, as opposed to the Adam optimizer or other gradient descent variants, resulted in models that better captured the high-frequency vibronic peaks of the target spectral density (SI Fig.~\ref{si:fig:lbfgs-vs-adam}). We trained models on 50, 100, 200, and 400 embedded EOM-CCSD energy gaps to construct the learning curve shown in Fig.~\ref{fig:summary_schematic}. In each case, we employed a validation set of 35 configurations for early stopping.

The second transfer learning step uses the weights from the trained hidden-solvent models to initialize indirect-solvent models that we train using the higher-level energy gaps. Here, we also used the L-BFGS optimizer in the same way we used it to optimize the hidden-solvent models' weights and used the same validation set for early stopping.

\section{Calculating spectra via the cumulant approximation}
\label{si:sec:cumulant-approach}
In this work, we calculated linear and two-dimensional electronic absorption spectra (2DES) via a cumulant expansion of the energy gap operator truncated at the 2nd order\cite{Mukamel1995}. Below we detail the working equations for calculating spectra using this approach and refer the reader to the corresponding references for further details.

This approach involves calculating the second order cumulant lineshape function $g_2(t)$\cite{Mukamel1995},
\begin{equation}
    \label{eqn:lineshape}
    g_2(t)=\frac{1}{\pi}\int_0^\infty \textrm{d}\omega\, \frac{\mathscr{J}(\omega)}{\omega^2}\left[\textrm{coth}\left(\frac{\beta \omega}{2}\right)\left[1-\cos{\omega t}\right] -\textrm{i}\left[\sin{\omega t}-\omega t\right] \right],
\end{equation}
where $\mathscr{J}(\omega)$ is the spectral density. The spectral density is a functional of the quantum time correlation function of the energy gap fluctuation operator $\delta U$, where $\delta U(\boldsymbol{\hat{q}})=H_e(\boldsymbol{\hat{q}})-H_g(\boldsymbol{\hat{q}})-\omega_\textrm{eg}^{\textrm{av}}$ is the energy gap operator, $H_g$ and $H_e$ are the nuclear Hamiltonians of the electronic ground- and excited-state potential energy surface, and $\omega_\textrm{eg}^{\textrm{av}}\equiv \langle U \rangle$ is the average energy gap value. In order to render this procedure more computationally tractable, we instead approximate the spectral density using the classical time correlation function of the energy gap fluctuations ($C_{\delta U}^{cl} (t)$) as follows,
\begin{equation}
    \label{eqn:spectral_density}
    \mathscr{J}(\omega)=\theta(\omega)\frac{\beta\omega}{2}\int_{-\infty}^\infty \textrm{d}t\,e^{\textrm{i}\omega t}C_{\delta U}^{\textrm{cl}}(t)e^{-|t|/\tau},
\end{equation}
where $\theta(\omega)$ is the Heaviside step function and the decaying exponential $e^{-|t|/\tau}$ is introduced to guarantee that the Fourier transform of $C^\textrm{cl}_{\delta U}(t)$ is well-behaved. For all calculations reported in this work, a decay constant of $\tau=500$~fs was used.

We calculate the linear absorption spectrum $\sigma(\omega)$ as,
\begin{equation}
    \label{eqn:linear_abs_spectrum}
    \sigma(\omega) = \alpha(\omega)\int_{-\infty}^\infty e^{\textrm{i}\omega t}\chi(t),
\end{equation}
where $\alpha(\omega)$ is a prefactor that for our calculations we treat as a constant and the linear response function $\chi(t)$ is approximated with the $g_2(t)$ cumulant lineshape function truncated to second order,
\begin{equation}
    \label{eqn:linear_response}
    \chi(t)=|\mu_\textrm{eg}|^2 e^{\textrm{i}\omega_\textrm{eg}^\textrm{av}t-g_2(t)},
\end{equation}
where $\mu_\textrm{eg}$ is the transition dipole moment between the electronic ground and excited state. For our calculations we apply the Condon approximation\cite{Condon1926theory,Condon1928nuclear}, i.e., the transition dipole moment $\mu_\textrm{eg}$ is assumed to be independent of the nuclear positions.

The 2DES spectra reported in this work are the purely absorptive spectra, obtained by adding the rephasing ($R_2$ and $R_3$) and non-rephasing ($R_1$ and $R_4$) contributions to the third-order response function and Fourier-transforming the $t_1$ and $t_3$ time variables. For a given delay time $t_2=t_\textrm{delay}$, the absorptive 2DES spectrum can then be expressed as\cite{Mukamel1995}:
\begin{equation}
\label{eqn:2DES}
\begin{split}
S_{\textrm{2DES}}(\omega_3,t_\textrm{delay},\omega_1)\propto \textrm{Re} \int_0^\infty\textrm{d}t_3\int_0^\infty\textrm{d}t_1 \times\\
\big[e^{\textrm{i}\omega_3t_3+\textrm{i}\omega_1t_1}
\left(R_1(t_3,t_\textrm{delay},t_1)+R_4(t_3,t_\textrm{delay},t_1)\right) \\
 + e^{-\textrm{i}\omega_3t_3+\textrm{i}\omega_1t_1}\left(R_2(t_3,t_\textrm{delay},t_1)+R_3(t_3,t_\textrm{delay},t_1)\right)\big],
\end{split}
\end{equation}
where the individual contributions to the third-order response are
\begin{equation}
    \label{eqn:third_order_response}
    \begin{split}
        R_1(t_3,t_2,t_1)=&|\mu_\textrm{eg}|^4 e^{-\textrm{i}\omega_\textrm{eg}^\textrm{av}(t_1 + t_3)} e^{-g_2(t_1)-g_2^*(t_2)-g_2^*(t_3)+g_2(t_1+t_2)+g^*_2(t_2+t_3)-g_2(t_1+t_2+t_3)},\\
        R_2(t_3,t_2,t_1)=&|\mu_\textrm{eg}|^4 e^{-\textrm{i}\omega_\textrm{eg}^\textrm{av}(t_3 - t_1)} e^{-g^*_2(t_1)+g_2(t_2)-g_2^*(t_3)-g^*_2(t_1+t_2)-g_2(t_2+t_3)+g^*_2(t_1+t_2+t_3)}, \\
        R_3(t_3,t_2,t_1)=&|\mu_\textrm{eg}|^4 e^{-\textrm{i}\omega_\textrm{eg}^\textrm{av}(t_3 - t_1)} e^{-g^*_2(t_1)-g_2(t_2)-g_2(t_3)-g^*_2(t_1+t_2)-g^*_2(t_2+t_3)+g^*_2(t_1+t_2+t_3)}, \\
        R_4(t_3,t_2,t_1)=&|\mu_\textrm{eg}|^4 e^{-\textrm{i}\omega_\textrm{eg}^\textrm{av}(t_1 + t_3)} e^{-g_2(t_1)-g_2(t_2)-g_2(t_3)+g_2(t_1+t_2)+g_2(t_2+t_3)-g_2(t_1+t_2+t_3)}.
    \end{split}
\end{equation}

\section{MD simulation details}
\label{si:sec:md-details}
The spectra we report here, and the underlying time correlation function of energy gap fluctuations, were calculated using a previously published set of 10 \textit{ab initio} molecular dynamics (AIMD) trajectories \cite{Zuehlsdorff2018unraveling} generated using the revPBE functional\cite{Perdew1996generalized,Zhang1998comment} with Grimme's D3 dispersion correction\cite{Grimme2010consistent}, that were each 15~ps in length with configurations saved at every 2~fs. This amounts to 75,000 MD configurations for which electronic excited-state energies need to be evaluated. In using our ML model as a proxy for computationally expensive EOM-CCSD evaluations, we save $\sim$30 million CPU core hours and thus enable a study that would otherwise be prohibitively expensive.

\section{Hysteresis effects in EOM-CCSD transfer learning models when not utilizing the hidden-solvent model}

\begin{figure*}[h]
    \begin{center}
        \includegraphics[width=0.5\textwidth]{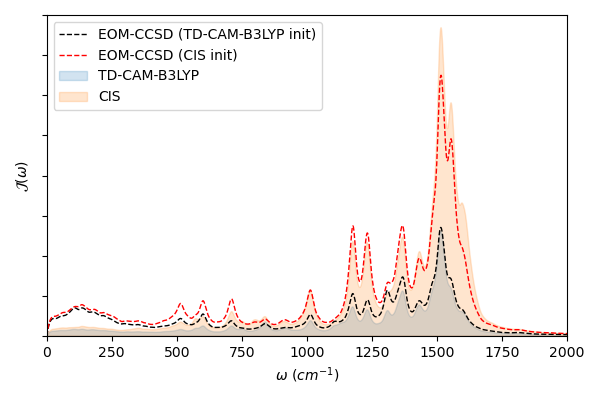}
    \end{center}
    \caption{Spectral densities obtained from ML models trained on 400 embedded EOM-CCSD energy gaps for the GFP chromophore in water using a transfer learning procedure where we do not train a hidden-solvent model and initial weights from either a gradient-trained TD-CAM-B3LYP (black) or CIS (red) ML model. The results obtained from our two transfer-learned embedded EOM-CCSD models are not consistent with each other and show hysteresis in the high-frequency peaks.}
	\label{si:fig:skip-hidsolv}
\end{figure*}

When we skipped training a hidden-solvent model we observed that the transfer learning models trained to embedded EOM-CCSD energy gaps displayed significant hysteresis in the high-frequency peaks of the spectral densities. SI Fig.~\ref{si:fig:skip-hidsolv} shows that when we initialized this alternative transfer learning procedure with TD-CAM-B3LYP that the intensities of the high-frequency parts are left unchanged. This was also the case when we initialized the procedure with CIS, training on the same set of 400 embedded EOM-CCSD energy gaps. From this test, we concluded that the training of the hidden-solvent model as an intermediate step in our transfer learning procedure was essential for properly correcting for the high-frequency intensities in the spectral density.

\section{Unshifted linear spectra and gas-phase spectral densities}
\begin{figure*}[h]
    \begin{center}
        \includegraphics[width=0.95\textwidth]{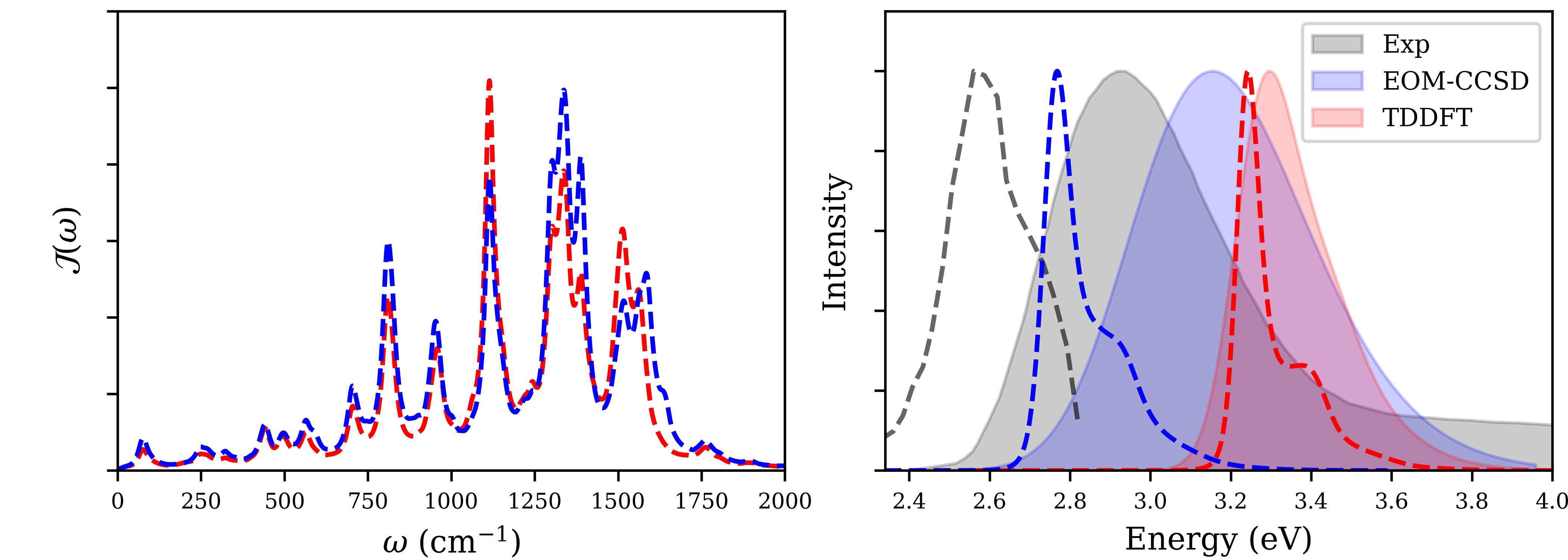}
    \end{center}
    \caption{(left) Spectral densities for the GFP chromophore in the gas phase obtained using our ML models trained to TDDFT and EOM-CCSD; (right) Unshifted linear spectra for the GFP chromophore in water (shaded) obtained using our ML models for TDDFT and EOM-CCSD showing that the EOM-CCSD spectrum's peak position is closer to that of the experiment\cite{Nielsen2001}. The unshifted gas-phase spectra (dashed lines) also highlight that solvation leads to a larger shift in peak positions for EOM-CCSD as compared to TDDFT.}
	\label{si:fig:linspec-unshifted}
\end{figure*}
Figure~\ref{fig:specdens_linspec} in the main paper shows our simulated linear electronic absorption spectra where we shifted the peak positions to match that of the experimental spectrum in order to compare the shapes of the spectra. In SI Figure~\ref{si:fig:linspec-unshifted}, we show the same simulated linear spectra but without shifting peak positions in order to highlight the difference in average predicted energy gaps between our TDDFT and EOM-CCSD ML models for the GFP chromophore in the gas-phase and in water. The linear spectrum for the solvated system obtained via EOM-CCSD has a peak position that is blue-shifted with respect to the experiment but closer than what we obtained via TDDFT. The blue-shift of the solvated EOM-CCSD spectrum might be partially attributed to the artificial blue-shift that the embedding procedure introduces, which we characterized by conducting analogous embedded TDDFT in DFT calculations and comparing results with a full TDDFT treatment of the system (SI Fig.~\ref{si:fig:eTDDFT-check}). Solvation also produces a markedly larger peak shift for EOM-CCSD as compared to TDDFT, which we see in comparing the respective gas-phase and solvated peak positions in SI Figure~\ref{si:fig:linspec-unshifted}.

\section{Low-frequency spectral density contribution to linear spectrum broadening}
\begin{figure*}[h]
    \begin{center}
        \includegraphics[width=0.95\textwidth]{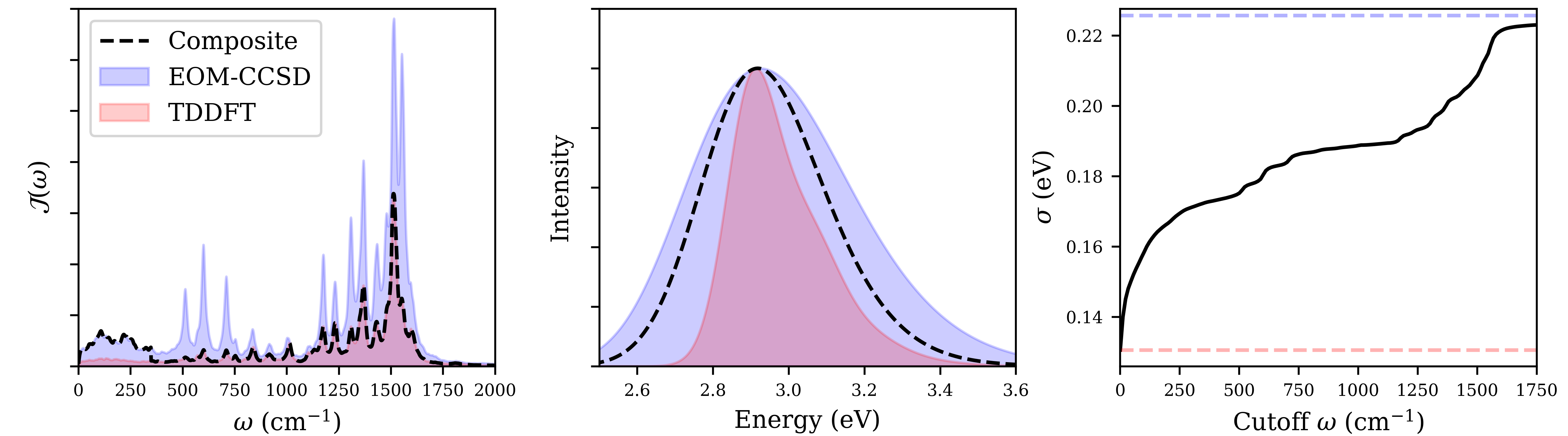}
    \end{center}
    \vspace{-4mm}
    \caption{A composite spectral density (left), using $<350$cm$^{-1}$ from EOM-CCSD and $>350$cm$^{-1}$ from TDDFT, and the corresponding linear spectrum (middle) are shown as dashed black lines. (right) A scan showing how the width of the spectrum, quantified by the square root of its second moment ($\sigma$), varies with the cutoff frequency employed. The red and blue dashed lines indicate the widths of the TDDFT and EOM-CCSD spectra, respectively.}
    \label{si:fig:specdens-stitch}
\end{figure*}
In SI Figure \ref{si:fig:specdens-stitch}, we build a composite spectral density composed of the low-frequency part of the EOM-CCSD spectral density ($\omega<350$~cm$^{-1}$) and the higher frequency part of the TDDFT spectral density ($\omega>350$~cm$^{-1}$) and show how the resulting linear spectrum obtained from this composite spectral density recaptures most of the width of the EOM-CCSD linear spectrum. Hence, we show that it is the difference in intensities at the lower frequencies of the spectral densities that predominantly leads to the EOM-CCSD spectrum being broader than the TDDFT spectrum. However, the difference in intensities at higher frequencies also contributes considerably to the difference in widths as well, as we see when we scan the cutoff frequency used to combine the spectral densities and plot the resulting widths of the composite spectra in the right panel of SI Figure \ref{si:fig:specdens-stitch}. In order to quantify the widths of the spectra, we calculate the square root of its second moment ($\sigma$). The widths of the TDDFT and EOM-CCSD spectra are 0.13~eV and 0.23~eV, while the width of the composite spectrum with a cutoff at 350~cm$^{-1}$ is 0.17~eV.

\begin{figure*}[h]
    \begin{center}
        \includegraphics[width=0.5\textwidth]{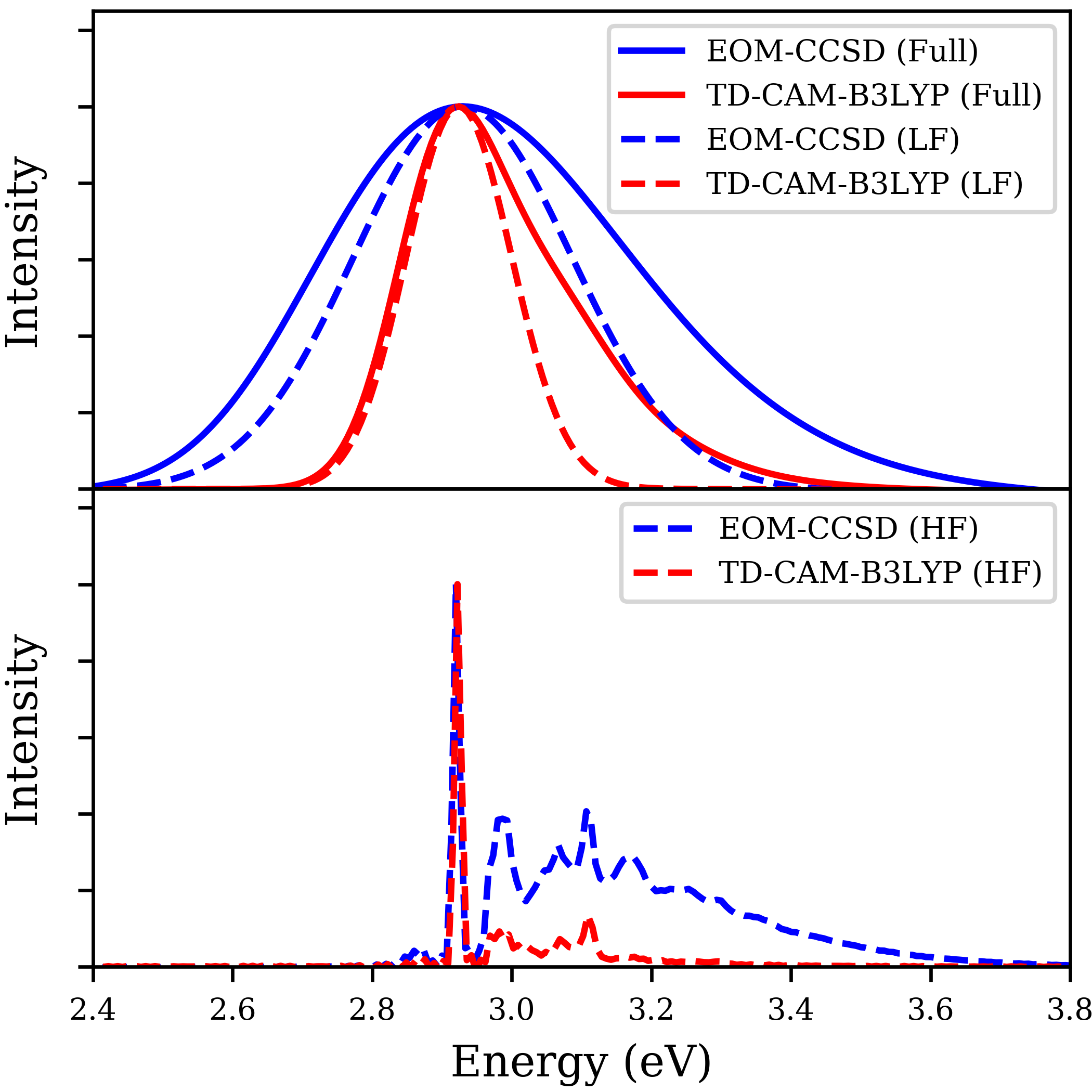}
    \end{center}
    \vspace{-4mm}
    \caption{(top) Linear absorption spectra obtained from using the low-frequency ($<$~350 cm$^{-1}$) part of the EOM-CCSD (blue) or TD-CAM-B3LYP (red) spectral densities as compared to the linear spectra obtained from the full spectral densities. (bottom) Linear spectra obtained from using the high-frequency ($>$~350 cm$^{-1}$) parts of the spectral densities.}
    \label{si:fig:lf-vs-hf}
\end{figure*}

Figure \ref{si:fig:lf-vs-hf} shows the result of separating the low- and high-frequency contributions of the spectral density on the resulting linear absorption spectra using the same 350~cm$^{-1}$ cutoff value. In the upper figure, the increase in the total spectral broadening is shown to be clearly dominated by the low-frequency region. There is also an increase in broadening due to increased vibronic coupling as seen in the lower figure, but the contribution is smaller in magnitude. 

\section{Average energy gaps for different hydrogen-bonding environments}
\label{si:sec:double-hb-envs}

\begin{table}[h]
    \centering
    \begin{tabular}{ | m{0.14\textwidth} || m{0.09\textwidth} | m{0.09\textwidth}| m{0.09\textwidth} | } 
        \hline
        TDDFT & 3 HB to C=O & 2 HB to C=O & 1 HB to C=O \\ \hline\hline
        4 HB to O$^{-}$ & 3.37 & 3.40 & 3.41 \\ \hline
        3 HB to O$^{-}$ & 3.34 & 3.37 & 3.39 \\ \hline
        2 HB to O$^{-}$ & 3.31 & 3.34 & 3.36 \\ \hline
    \end{tabular}
    \begin{tabular}{ | m{0.14\textwidth} || m{0.09\textwidth} | m{0.09\textwidth}| m{0.09\textwidth} | } 
        \hline
        EOM-CCSD & 3 HB to C=O & 2 HB to C=O & 1 HB to C=O \\ \hline\hline
        4 HB to O$^{-}$ & 3.19 & 3.26 & 3.35 \\ \hline
        3 HB to O$^{-}$ & 3.12 & 3.20 & 3.28 \\ \hline
        2 HB to O$^{-}$ & 3.05 & 3.13 & 3.19 \\ \hline
    \end{tabular}
    \caption{Average electronic energy gaps in eV predicted by our TDDFT (left) and EOM-CCSD (right) ML model for the GFP chromophore in water depending on how many hydrogen bonds are formed with the O$^{-}$ and C=O groups on the chromophore}
    \label{si:tab:tddft-hb-avg-egaps}
\end{table}

SI Table~\ref{si:tab:tddft-hb-avg-egaps} shows the average electronic energy gaps for the GFP chromophore in water for different hydrogen bonding environments using both our TDDFT and EOM-CCSD ML models for the solvated energy gaps. We examine in particular the different combinations of hydrogen bonds formed with the C=O and O$^{-}$ groups of the chromophore, where we identify a hydrogen bond when the O$_\text{w}\cdots$O$_\text{C}$ distance is below 2.5 \AA (see SI Sec.~\ref{si:sec:hb-criteria}). We find that the smallest average energy gap results from configurations where the chromophore forms 2 and 3 hydrogen bonds to the C=O and O$^{-}$ groups, respectively. On the other hand, the largest average energy gap is observed when the chromophore forms 1 and 4 hydrogen bonds to the C=O and O$^{-}$ groups, respectively. The difference in magnitudes between these two extremes is considerably larger for EOM-CCSD (3.31 vs. 3.41~eV) as compared to TDDFT (3.05 vs. 3.35~eV).

\section{Quantifying broadening of energy gap distributions due to solvation}
\label{si:sec:quantify_solv_broadening}
\begin{figure*}[h]
    \begin{center}
        \includegraphics[width=0.85\textwidth]{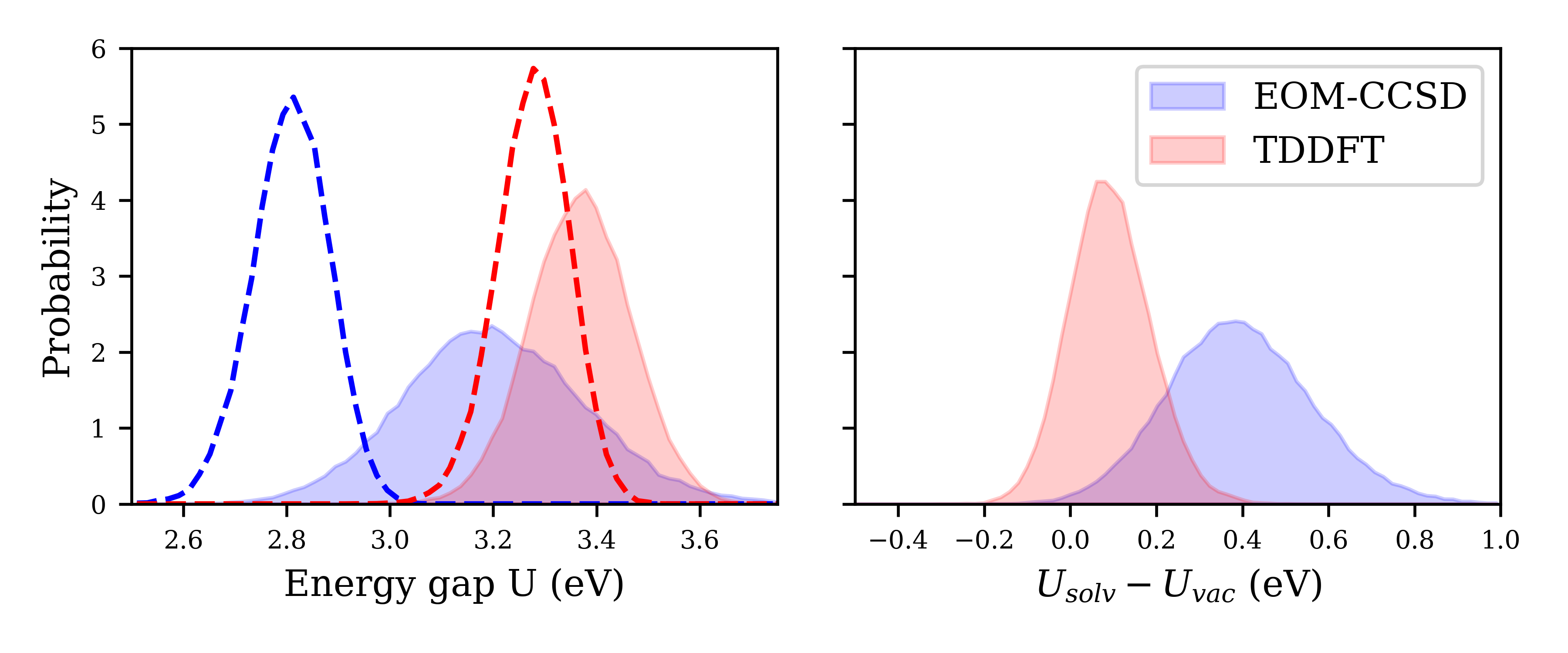}
    \end{center}
    \caption{(left) Distributions for the energy gaps of the solvated chromophore ($U_{solv}$, shaded) and for the same chromophore configuration but in vacuum ($U_{vac}$, dashed lines). The distributions were obtained from applying our solvated and gas-phase ML models trained to TDDFT (red) and EOM-CCSD (blue) to AIMD trajectories of the GFP chromophore in water\cite{Zuehlsdorff2018unraveling}. (right) Distributions of energy gap shifts due to solvent, $\Delta U = U_{solv}-U_{vac}$, for EOM-CCSD and TDDFT.}
	\label{si:fig:egap-histograms}
\end{figure*}
We applied our ML models for the solvated and gas-phase GFP chromophore to AIMD trajectories of the GFP chromophore in water\cite{Zuehlsdorff2018unraveling} in order to compare the extent to which solvation as modeled via EOM-CCSD and TDDFT broadens the energy gap distribution for the chromophore. To conduct this analysis, we calculate a change in the energy gap due to solvation, $\Delta U = U_{solv}-U_{vac}$, for each solvated chromophore configuration from our AIMD trajectories. We obtain $U_{solv}$ and $U_{vac}$ using our solvated and gas-phase ML models, respectively, and in the case of the latter we only use the atomic positions of the chromophore (i.e., ignore solvent) in our energy gap prediction. The standard deviations for the EOM-CCSD and TDDFT $\Delta U$ distributions are $\sigma_{\Delta U,\text{EOM-CCSD}}=0.168$~eV and $\sigma_{\Delta U,\text{TD-CAM-B3LYP}}=0.095$~eV.

\section{Hydrogen bonding environment criteria}
\label{si:sec:hb-criteria}
\begin{figure*}[h]
    \begin{center}
        \includegraphics[width=0.75\textwidth]{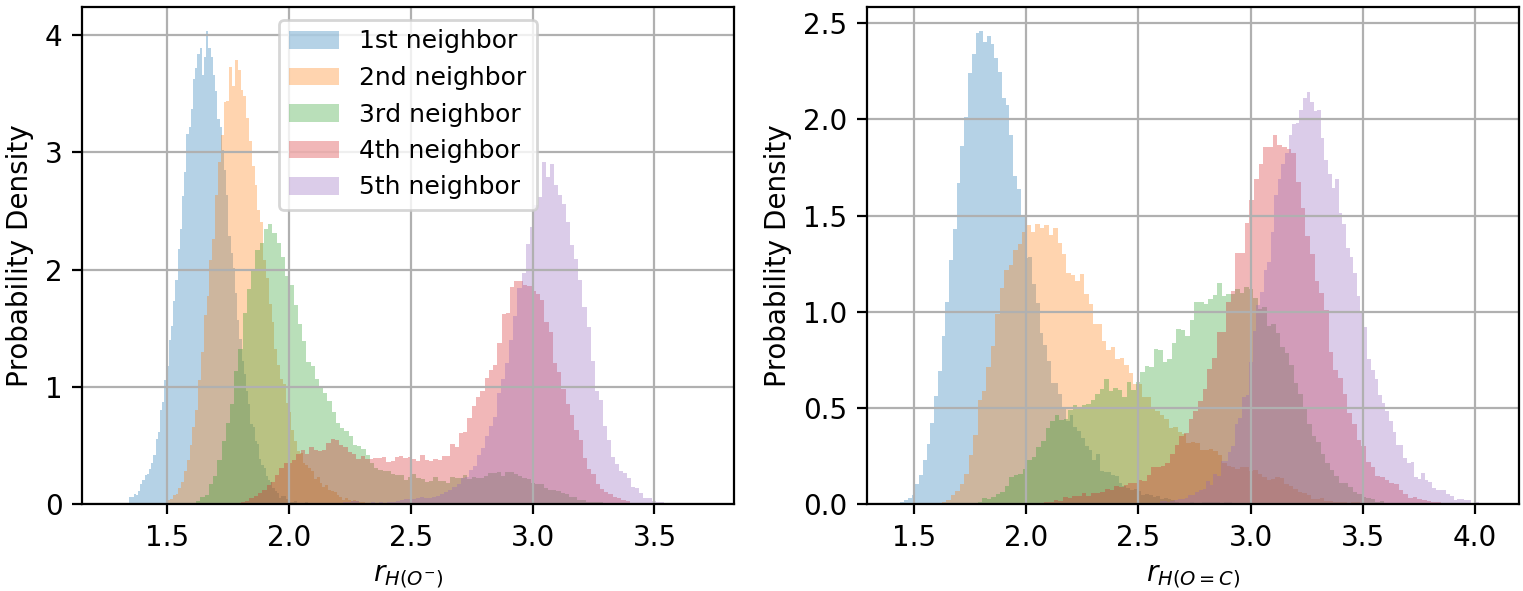}
    \end{center}
    \caption{Nearest-neighbor hydrogen radial density distributions around the O$^{-}$ (left) and C=O (right) groups on the GFP chromophore. Radial distances are given in Angstroms from the oxygen atom of the specific group.}
	\label{si:fig:hb-rdfs}
\end{figure*}

For the GFP chromophore in water, we used a 2.5~\AA\ cutoff for the O$_\text{C}\cdots$O$_\text{w}$  distance to assign how many hydrogen bonding interactions there were to the O$^{-}$ or C=O moiety on the GFP chromophore. We decided on this cutoff after examining the the nearest neighbor radial density distributions shown in SI Fig.~\ref{si:fig:hb-rdfs}. From these distributions, 2.5~\AA\ presents itself as a natural cutoff for assigning whether the O$^{-}$ (C=O) moiety has 2, 3, or 4 (1, 2, 3, or 4) hydrogen bonding interactions.

\section{TRPMD results}
\begin{figure*}[h]
    \begin{center}
        \includegraphics[width=0.85\textwidth]{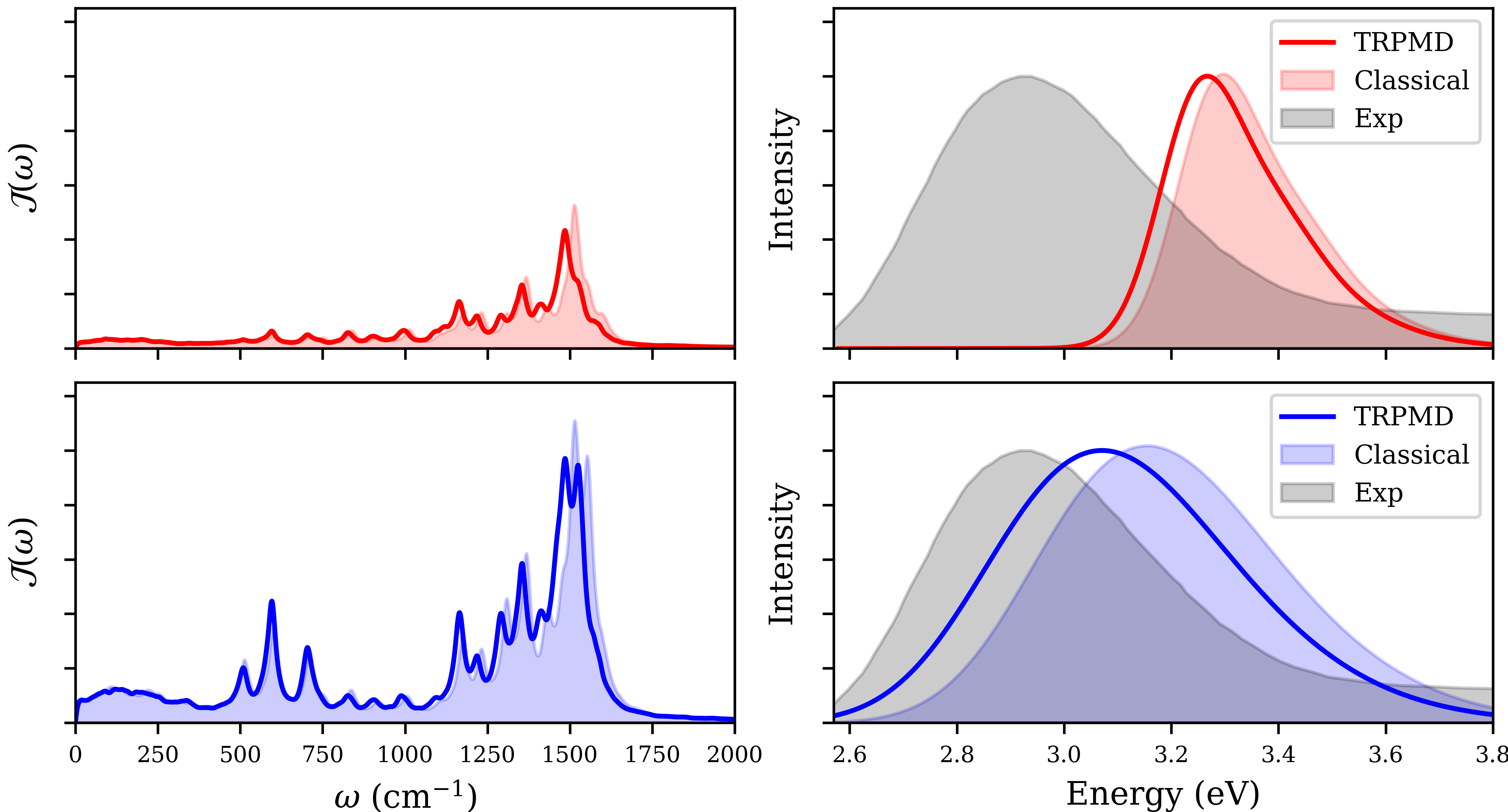}
    \end{center}
    \caption{Comparing spectral densities (left) and linear spectra (right) for the GFP chromophore in water obtained via classical MD versus TRPMD using our ML models for TD-CAM-B3LYP (top row) and EOM-CCSD (bottom row).}
	\label{si:fig:rpmd}
\end{figure*}

We calculate the TRPMD spectra via the 2nd-order cumulant approach. The bottom row of SI Fig.~\ref{si:fig:rpmd} compares the spectral densities and linear spectra we obtain from using classical MD or TRPMD and our ML model for the embedded EOM-CCSD energy gaps for the solvated GFP chromophore. The spectral densities obtained from classical MD and TRPMD are quite similar aside from a slight red shift of the vibronic peaks that seems more pronounced for the higher-frequency peaks. Consequently we see that the corresponding linear spectra are also similar in shape, although the TRPMD spectrum is red-shifted from 3.16~eV to 3.07~eV and thus closer to the peak position of the experimental spectrum at 2.94~eV. From SI Fig.~\ref{si:fig:rpmd} that this red-shift is smaller in magnitude for TD-CAM-B3LYP, which shifts from 3.30~eV to 3.27~eV. 

\begin{figure*}[h]
    \begin{center}
        \includegraphics[width=0.85\textwidth]{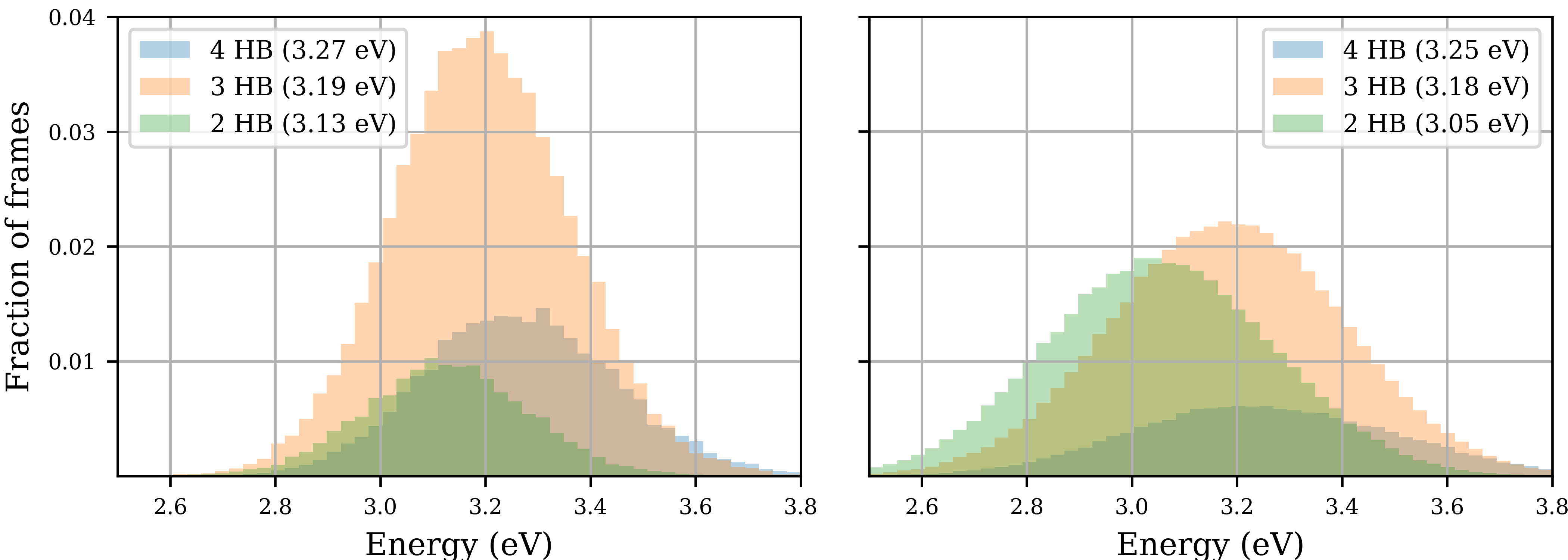}
    \end{center}
    \caption{Comparison of the ML EOM-CCSD energy gap distributions we obtain from  classical (left) and TRPMD (right) trajectories when the phenolate oxygen is involved in 2, 3, or 4 hydrogen bonds. Average energy gaps for each separate distribution are provided in the legends.}
	\label{si:fig:hb-rpmd}
\end{figure*}

SI Figure \ref{si:fig:hb-rpmd} shows a comparison of the the hydrogen-bond dependent EOM-CCSD energy gap distributions we obtain from  classical (left) and TRPMD (right) trajectories, focusing on the number of hydrogen bonds made with the phenolate oxygen on the GFP chromophore. Note that the classical MD results are the same as what were presented in Figure \ref{fig:hb}, however here we normalize the distributions such that relative intensities between distributions are comparable whereas in the main text each conditional distribution is normalized such that their integral is 1. This comparison highlights that the red shift we observe in our linear absorption spectrum when incorporating nuclear quantum effects via TRPMD arises from the sampling of more configurations of the chromophore where two hydrogen bonds are formed with the phenolate oxygen.

\clearpage
\bibliography{bibliography}